\documentclass[aps,showpacs,amsmath,amssymb]{revtex4}
\usepackage{feynmp}
\usepackage{supercite}
\newcommand{\setval}{\fmfset{wiggly_len}{1.5mm}\fmfset{arrow_len}{1.5mm}\fmfset{arrow_ang}{13}\fmfset{dash_len}{1.5mm}\fmfpen{0.125mm}\fmfset{dot_size}{1thick}}
\newcommand{\dphi}[3]{\frac{\delta #1}{\delta
\setlength{\unitlength}{1mm}
\parbox{9mm}{\centerline{
\begin{fmfgraph*}(5,3)
\setval
\fmfleft{v1}
\fmfright{v2}
\fmf{fermion}{v2,v1}
\fmfv{decor.size=0,label=${\scriptstyle #2}$,l.dist=0.5mm}{v1}
\fmfv{decor.size=0,label=${\scriptstyle #3}$,l.dist=0.5mm}{v2}
\end{fmfgraph*}
}}}}
\newcommand{\ddphi}[1]{\frac{\delta^2 #1}{\delta
               \parbox{9mm}{\centerline{
\begin{fmfgraph*}(5,3)
\setval
\fmfleft{v1}
\fmfright{v2}
            \fmf{fermion}{v2,v1}
\fmfv{decor.size=0,label=${\scriptstyle 1}$,l.dist=0.5mm}{v1}
\fmfv{decor.size=0,label=${\scriptstyle 2}$,l.dist=0.5mm}{v2}
\end{fmfgraph*}
}}\,\delta
\parbox{9mm}{\centerline{
\begin{fmfgraph*}(5,3)
\setval
\fmfleft{v1}
\fmfright{v2}
\fmf{fermion}{v2,v1}
\fmfv{decor.size=0,label=${\scriptstyle 3}$,l.dist=0.5mm}{v1}
\fmfv{decor.size=0,label=${\scriptstyle 4}$,l.dist=0.5mm}{v2}
\end{fmfgraph*}
}}
}}
\begin{document}
\begin{fmffile}{hugo1}
\title{Many-Body Vacuum Diagrams and Their Recursive Graphical Construction}
\author{Axel Pelster and Konstantin Glaum}
\affiliation{Institut f\"ur Theoretische Physik,
Freie Universit\"at Berlin, Arnimallee 14, 14195 Berlin, Germany\\
{\tt pelster@physik.fu-berlin.de, glaum@physik.fu-berlin.de}\\$\mbox{}$ }
\date{\today}
\begin{abstract}
The grand-canonical potential of many-body physics can be considered as a functional
of the interaction-free correlation function. As such it obeys a 
nonlinear functional differential equation which can be turned into a 
recursion relation. This is solved graphically
order by order in the two-particle interaction to find all 
connected vacuum diagrams with their proper weights. 
As a special case, the procedure is applied to generate the Hugenholtz diagrams for a weakly
interacting Bose gas.
\end{abstract}
\pacs{24.10.Cn}
\maketitle
\section{Introduction}
A systematic and physical approach to construct all Feynman diagrams of a quantum field theory together with their weights was
proposed some time ago by Kleinert \cite{Kleinert1}. It is based on a functional differential equation for the vacuum energy which
involves functional derivatives with respect to free propagators and interactions. Solving the functional differential equation
by a recursive graphical procedure leads to all vacuum diagrams. In a subsequent step, the diagrams of $n$-point functions are found
graphically by amputating lines and vertices in the vacuum diagrams. Recently, this approach was used to systematically
generate all connected and one-particle irreducible Feynman diagrams of the euclidean multicomponent $\phi^4$-theory both in the
disordered symmetric phase \cite{SYM} and in the ordered, spontaneously broken-symmetry phase \cite{ASYM1,ASYM2}. The
approach was also applied to QED \cite{QED} and scalar QED \cite{GINZ}. Furthermore, a modification of this program led to a
closure of the infinite hierarchy of Schwinger-Dyson equations for the $n$-point functions in a certain functional sense.
Both in $\phi^4$-theory \cite{Glaum,FEST} and in QED \cite{SDQED1} this allowed us to construct most directly all
diagrams for $n$-point functions which are relevant for the renormalization of the theory.\\

Here we shall use this approach to generate iteratively the vacuum diagrams of many-body theory. We start with briefly
reviewing some relevant graphical operations in Section \ref{MBP}. They are applied in Section \ref{per} to determine the 
grand-canonical potential as a power series of the two-particle interaction. In Section \ref{graph} we derive the functional
differential equation for the grand-canonical potential and show how to solve it graphically order by order. Finally,
Section \ref{HuGo} discusses as a special case the resulting recursive graphical construction of the Hugenholtz diagrams
for a weakly interacting Bose gas.
\section{Many-Body Physics} \label{MBP}
The thermal fluctuations of a quantum many-particle system are controlled by the action functional
\begin{eqnarray}
\label{AF}
{\cal A} [ \psi^* , \psi ] = \int_{12} G^{-1}_{12} \psi^*_1 \psi_2 
+ \frac{1}{2} \int_{1234} V_{1234}^{({\rm int})}  \psi_1^* \psi_2 \psi_3^* \psi_4 \,,
\end{eqnarray}
where $\psi^*,\psi$ denote complex fields for bosons and Grassmann fields for fermions.
In this short-hand notation, the number indices of
the fields $\psi^*,\psi$, the bilocal kernel $G^{-1}$, and the two-particle 
interaction $V^{({\rm int})}$ are abbreviations for the spatial coordinates and the imaginary time, i.e.,
\begin{eqnarray}
1 \equiv \{ {\bf x}_1 , \tau_1 \} \, , \,\, 
\int_1 \equiv \int d^3 x_1  \int_0^{\beta} d \tau_1 \, , \,\,
\psi_1^* \equiv \psi^* ( {\bf x}_1 , \tau_1 ) \, , \,\,
\psi_1 \equiv \psi ( {\bf x}_1 , \tau_1 ) \, ,  \nonumber\\
G^{-1}_{12} \equiv G^{-1} ( {\bf x}_1 , \tau_1 ; {\bf x}_2 , \tau_2 ) \, , \,\, 
V^{({\rm int})}_{1234} \equiv V^{({\rm int})} ( {\bf x}_1 , \tau_1 ; {\bf x}_2 , \tau_2 ; {\bf x}_3 , \tau_3 ; {\bf x}_4 , \tau_4 ) \, . 
\end{eqnarray}
Both for bosons and for fermions, the action functional (\ref{AF}) is specified by the kernel
\begin{eqnarray}
\label{PH}
G^{-1} ( {\bf x}_1 , \tau_1 ; {\bf x}_2 , \tau_2 ) = \delta ( {\bf x}_1 - {\bf x}_2 ) \, \delta (\tau_1 - \tau_2)
\left[ \frac{\partial}{\partial \tau_2} - \frac{1}{2 M} \, \Delta_2 + V ({\bf x}_2) - \mu \right] \, , 
\end{eqnarray}
where $M$ denotes the mass of the particles and $V$ represents the one-particle potential.
Furthermore, the two-particle interaction $V^{({\rm int})}$ is a functional tensor with the symmetry
\begin{eqnarray}
\label{SYM}
V^{({\rm int})}_{1234} = V^{({\rm int})}_{3412}  \, . 
\end{eqnarray}
In this paper we shall leave the kernel $G^{-1}$ in the action functional
(\ref{AF}) completely general, and insert
the physical value (\ref{PH}) only at the end. By doing so, we regard the action functional (\ref{AF})
as well as all local and global statistical quantities derived from (\ref{AF}) as
functionals of the bilocal kernel $G^{-1}$. In particular, we are interested in studying the 
partition function
\begin{eqnarray}
\label{PF}
Z = \int {\cal D} \psi^* \int {\cal D} \psi \,\, e^{- {\cal A} [ \psi^*,\psi ]} \, ,
\end{eqnarray}
where the functional integral is performed with respect to those fields $\psi^*,\psi$ which are periodic or antiperiodic 
in the imaginary time $\tau$,
depending on whether we treat bosons or fermions.
By expanding the functional integral (\ref{PF}) in powers of the two-particle interaction $V^{({\rm int})}$, the expansion
coefficients of the generating functional consist of interaction-free expectation values. These
are evaluated with the help of Wick's rule as a sum of Feynman integrals, which are pictured as diagrams constructed from
lines and vertices. 
The interaction-free correlation function $G$, which is the functional inverse of the kernel $G^{-1}$ 
in the action functional (\ref{AF})
\begin{eqnarray}
\label{FP}
\int_{2} G_{12} \, G^{-1}_{23} = \delta_{13} \, ,
\end{eqnarray}
is represented by a straight line with an arrow 
\begin{eqnarray}
\label{PRO}
G_{12} \hspace*{2mm}  \equiv \hspace*{4mm} 
\setlength{\unitlength}{1mm}
\parbox{9mm}{\centerline{
\begin{fmfgraph*}(7,3)
\setval
\fmfleft{v1}
\fmfright{v2}
\fmf{electron}{v2,v1}
\fmflabel{${\scriptstyle 1}$}{v1}
\fmflabel{${\scriptstyle 2}$}{v2}
\end{fmfgraph*}
}} 
\hspace*{4mm} \, ,
\end{eqnarray}
and the two-particle interaction $V^{({\rm int})}$ is pictured as two vertices which are connected by a dashed line 
\begin{eqnarray}
\label{VE}
- V_{1234}^{({\rm int})} \hspace*{2mm}  \equiv \hspace*{4mm} 
\setlength{\unitlength}{1mm}
\parbox{12mm}{\centerline{
\begin{fmfgraph*}(10,5)
\setval
\fmfforce{0w,0h}{i1}
\fmfforce{0w,1h}{i2}
\fmfforce{1w,0h}{o1}
\fmfforce{1w,1h}{o2}
\fmfforce{3/10w,0.5h}{v1}
\fmfforce{7/10w,0.5h}{v2}
\fmfforce{3/10w,1/10h}{v3}
\fmfforce{7/10w,1/10h}{v4}
\fmf{dashes}{v1,v2}
\fmf{electron}{v1,i1}
\fmf{electron}{i2,v1}
\fmf{electron}{o1,v2}
\fmf{electron}{v2,o2}
\fmfdot{v1,v2}
\fmfv{decor.size=0, label=${\scriptstyle 1}$, l.dist=1mm, l.angle=180}{i1}
\fmfv{decor.size=0, label=${\scriptstyle 2}$, l.dist=1mm, l.angle=180}{i2}
\fmfv{decor.size=0, label=${\scriptstyle 3}$, l.dist=1mm, l.angle=0}{o2}
\fmfv{decor.size=0, label=${\scriptstyle 4}$, l.dist=1mm, l.angle=0}{o1}
\end{fmfgraph*}}} 
\hspace*{2mm} \, .
\end{eqnarray}
The graphical elements (\ref{PRO}) and (\ref{VE}) are combined by an integral which graphically corresponds to
attaching a line to a vertex as, for instance,
\begin{eqnarray}
  - \int_4 V_{1234}^{({\rm int})} \hspace*{1mm} G_{45}
  \hspace*{3mm} \equiv  \hspace*{3mm}
  \setlength{\unitlength}{1mm}
\parbox{12mm}{\centerline{
\begin{fmfgraph*}(10,5)
\setval
\fmfforce{0w,0h}{i1}
\fmfforce{0w,1h}{i2}
\fmfforce{1w,0h}{o1}
\fmfforce{1w,1h}{o2}
\fmfforce{3/10w,0.5h}{v1}
\fmfforce{7/10w,0.5h}{v2}
\fmfforce{3/10w,1/10h}{v3}
\fmfforce{7/10w,1/10h}{v4}
\fmf{dashes}{v1,v2}
\fmf{electron}{v1,i1}
\fmf{electron}{i2,v1}
\fmf{electron}{o1,v2}
\fmf{electron}{v2,o2}
\fmfdot{v1,v2}
\fmfv{decor.size=0, label=${\scriptstyle 1}$, l.dist=1mm, l.angle=180}{i1}
\fmfv{decor.size=0, label=${\scriptstyle 2}$, l.dist=1mm, l.angle=180}{i2}
\fmfv{decor.size=0, label=${\scriptstyle 3}$, l.dist=1mm, l.angle=0}{o2}
\fmfv{decor.size=0, label=${\scriptstyle 4}$, l.dist=1mm, l.angle=0}{o1}
\end{fmfgraph*}}} 
  \hspace*{5mm}
  \parbox{9mm}{\centerline{
  \begin{fmfgraph*}(6,7.5)
  \setval
  \fmfforce{0w,0h}{v1}
  \fmfforce{1w,0h}{v2}
  \fmf{fermion,width=0.2mm}{v2,v1}
  \fmfv{decor,size=0, label=${\scriptstyle 4}$, l.dist=1mm, l.angle=-180}{v1}
  \fmfv{decor,size=0, label=${\scriptstyle 5}$, l.dist=1mm, l.angle=0}{v2}
  \end{fmfgraph*} } } 
  \hspace*{3mm} \equiv \hspace*{3mm}  
   \setlength{\unitlength}{1mm}
\parbox{12mm}{\centerline{
\begin{fmfgraph*}(10,5)
\setval
\fmfforce{0w,0h}{i1}
\fmfforce{0w,1h}{i2}
\fmfforce{1w,0h}{o1}
\fmfforce{1w,1h}{o2}
\fmfforce{3/10w,0.5h}{v1}
\fmfforce{7/10w,0.5h}{v2}
\fmfforce{3/10w,1/10h}{v3}
\fmfforce{7/10w,1/10h}{v4}
\fmfforce{14/10w,-2/10h}{v5}
\fmf{dashes}{v1,v2}
\fmf{electron}{v1,i1}
\fmf{electron}{i2,v1}
\fmf{electron}{v2,o2}
\fmf{electron,left=0.3}{v5,v2}
\fmfdot{v1,v2}
\fmfv{decor.size=0, label=${\scriptstyle 1}$, l.dist=1mm, l.angle=180}{i1}
\fmfv{decor.size=0, label=${\scriptstyle 2}$, l.dist=1mm, l.angle=180}{i2}
\fmfv{decor.size=0, label=${\scriptstyle 3}$, l.dist=1mm, l.angle=0}{o2}
\fmfv{decor.size=0, label=${\scriptstyle 5}$, l.dist=1mm, l.angle=0}{v5}
\end{fmfgraph*}}} 
\hspace*{10mm} .
\label{5}
\end{eqnarray}
In this paper we generate the subset of connected Feynman diagrams contributing to the generating functional (\ref{PF})
together with their weights. To this end we
introduce the functional derivative with respect to the interaction-free correlation function $G$ according to
\begin{eqnarray}
\label{DR2}
\frac{\delta G_{12}}{\delta G_{34}} =  \delta_{13} \, \delta_{42}  \, .
\end{eqnarray}
Such functional derivatives are represented graphically by removing a line
of a Feynman diagram in all possible ways  \cite{Kleinert1,SYM,ASYM1,ASYM2,FEST,Glaum,QED,SDQED1,GINZ}. 
For practical purposes it is convenient to use also
functional derivatives with respect to the bilocal kernel 
$G^{-1}$ whose basic rule reads
\begin{eqnarray}
\label{DR1}
\frac{\delta G^{-1}_{12}}{\delta G^{-1}_{34}} =  
\delta_{13} \delta_{42} \, .
\end{eqnarray}
As has been elaborated in detail in Refs. \cite{Kleinert1,SYM,ASYM1,ASYM2,FEST,Glaum,QED,SDQED1,GINZ}, such a
functional derivative is represented by a graphical operation which
cuts a line of a Feynman diagram in all possible ways. 
Indeed, from the identity (\ref{FP}) and the functional product rule, 
we find the effect of this derivative on the interaction-free correlation function
\begin{eqnarray}
\label{ACT}
- \frac{\delta G_{12}}{\delta G^{-1}_{34}} = G_{13} G_{42} \, ,
\end{eqnarray}
which has the graphical representation
\begin{eqnarray}
- \, \frac{\delta}{\delta G^{-1}_{34}} \hspace*{4mm}
\setlength{\unitlength}{1mm}
\parbox{9mm}{\centerline{
\begin{fmfgraph*}(7,3)
\setval
\fmfleft{v1}
\fmfright{v2}
\fmf{electron}{v2,v1}
\fmflabel{${\scriptstyle 1}$}{v1}
\fmflabel{${\scriptstyle 2}$}{v2}
\end{fmfgraph*}
}} 
\hspace*{4mm} = \hspace*{4mm}
\setlength{\unitlength}{1mm}
\parbox{9mm}{\centerline{
\begin{fmfgraph*}(7,3)
\setval
\fmfleft{v1}
\fmfright{v2}
\fmf{electron}{v2,v1}
\fmflabel{${\scriptstyle 1}$}{v1}
\fmflabel{${\scriptstyle 3}$}{v2}
\end{fmfgraph*}
}} 
\hspace*{8mm}
\setlength{\unitlength}{1mm}
\parbox{9mm}{\centerline{
\begin{fmfgraph*}(7,3)
\setval
\fmfleft{v1}
\fmfright{v2}
\fmf{electron}{v2,v1}
\fmflabel{${\scriptstyle 4}$}{v1}
\fmflabel{${\scriptstyle 2}$}{v2}
\end{fmfgraph*}
}} \hspace*{4mm} .
\end{eqnarray}
The functional derivatives with respect to the correlation function $G$ and the kernel $G^{-1}$ are related via
the functional chain rule
\begin{eqnarray}
\frac{\delta}{\delta G^{-1}_{12}} \hspace*{1mm} = - \hspace*{1mm} \int_{34} G_{31} G_{24}
\hspace*{1mm} \frac{\delta}{\delta G_{34}}  \hspace*{3mm} .
\label{CHAIN1}
\end{eqnarray}
Thus cutting a line is equivalent to amputating a line 
and adding two lines to the vertices to which the original line was connected. 
\section{Perturbation Theory} \label{per}
Field theoretic perturbation expressions are usually derived by  
introducing external currents $j^*,j$ into the action functional (\ref{AF})
which are linearly coupled to the fields $\psi^*,\psi$ \cite{Negele,Abrikosov,Fetter,Huang,Mahan,Gross}. 
Thus the partition function
(\ref{PF}) becomes in the presence of $j^*,j$ the generating functional $Z [ j^*,j ]$
which allows us to find all interaction-free
$n$-point functions from functional derivatives with respect to these
external currents $j^*,j$. 
To calculate these, 
it is possible to 
substitute two functional derivatives with respect to the currents $j^*,j$
by one functional derivative
with respect to the kernel $G^{-1}$. This reduces 
the number of functional derivatives in each order 
of perturbation theory by one half and has the additional advantage that
the introduction of the currents $j^*,j$ becomes superfluous.
\subsection{Current Approach}
Recall briefly the standard perturbative treatment, in which
the action functional
(\ref{AF}) is artificially extended by a source term
\begin{eqnarray}
{\cal A} [ \psi^*,\psi ; j^* , j ] = {\cal A} [ \psi^*,\psi ] - \int_1 \left( j_1^* \, \psi_1 + \psi^*_1 \,j_1\right) \, ,
\end{eqnarray}
where $j^*,j$ denote complex current fields for bosons and Grassmann current fields for fermions. 
The functional integral for the generating functional
\begin{eqnarray}
\label{GF}
Z [j^*,j]= \int {\cal D} \psi^*  \int {\cal D} \psi \,\, e^{- {\cal A} [ \psi^*,\psi ; j^* , j ]} 
\end{eqnarray}
is first explicitly calculated for a vanishing two-particle interaction $V^{({\rm int})}$, yielding
\begin{eqnarray}
\label{GFF}
Z^{(0)} [ j^*,j ] = \exp \left( \mp \, \mbox{Tr} \ln G^{-1} + 
\int_{12} \, G_{12} \, j_1^* j_2 \right) \, ,
\end{eqnarray}
where the trace of the logarithm of the kernel is defined by the series
\cite[p.~16]{Kleinert4}
\begin{eqnarray}
\label{LOG}
\mbox{Tr} \ln G^{-1} = \sum_{n = 1}^{\infty} \frac{(-1)^{n + 1}}{n}
\int_{1 \ldots n} \left( G^{-1}_{12} - \delta_{12} \right) \cdots
\left( G^{-1}_{n1} - \delta_{n1} \right) \, .
\end{eqnarray}
If the two-particle interaction $V^{({\rm int})}$ does not vanish, one expands the generating
functional $Z [ j^*,j ]$ in powers of $V^{({\rm int})}$,
and reexpresses the resulting
powers of the fields
within the functional integral (\ref{GF}) as functional
derivatives with respect to the currents $j^*,j$. The original partition function
(\ref{PF}) can thus be obtained from the interaction-free generating functional
(\ref{GFF}) by the formula
\begin{eqnarray}
\label{CP}
Z = \left.\exp \left\{ - \frac{1}{2} \, \int_{1234} V_{1234}^{({\rm int})} \, 
\frac{\delta^4}{\delta j_1 \delta j_2^* \delta j_3 \delta j_4^*} \right\}
Z^{(0)} [ j^*, j ]\right|_{j^*=0,j=0} \, . 
\end{eqnarray}
Expanding the exponential in a power series, we arrive at the perturbation expansion
\begin{eqnarray}
Z &=& \left\{ 1 + \frac{-1}{2} \, \int_{1234} V_{1234}^{({\rm int})} \, 
\frac{\delta^4}{\delta j_1 \delta j_2^* \delta j_3 \delta j_4^*} +  \frac{1}{2} \left(  \frac{-1}{2} \right)^2 \int_{12345678}
V_{1234}^{({\rm int})} V_{5678}^{({\rm int})} 
\right.\nonumber \\
&& \left. \left. \times
\frac{\delta^8}{\delta j_1 \delta j_2^* \delta j_3 \delta j_4^*
\delta j_5 \delta j_6^* \delta j_7 \delta j_8^*} + \ldots \right\}
Z^{(0)} [ j^*, j ]\right|_{j^*=0,j=0} \, , \label{EJ}
\end{eqnarray}
in which the $p$th order contribution
for the partition function requires the evaluation of $4p$ functional
derivatives with respect to the currents $j^*,j$.
\subsection{Kernel Approach}
The derivation of the perturbation expansion simplifies,
if we abandon introducing the currents $j^*,j$ and
use instead functional derivatives
with respect to the kernel $G^{-1}$ in the action functional (\ref{AF}).
In this way we obtain for the partition function (\ref{PF}) instead of
the previous expression (\ref{CP}) the result
\begin{eqnarray}
\label{CPN}
Z = \exp \left\{ - \frac{1}{2} \, \int_{1234} V_{1234}^{({\rm int})} \, 
\frac{\delta^2}{\delta G^{-1}_{12} \delta G^{-1}_{34}} \right\}
e^{\Omega^{(0)}} \, ,
\end{eqnarray}
where the zeroth order of the grand-canonical potential $\Omega^{(0)}= \ln Z^{(0)}$ 
has the diagrammatic representation
\begin{eqnarray}
\label{FPF}
\Omega^{(0)} = \mp \, \mbox{Tr} \ln G^{-1} \equiv \mp \,\,\,
\setlength{\unitlength}{1mm}
\parbox{8mm}{\centerline{
\begin{fmfgraph}(5,5)
\setval
\fmfi{fermion}{reverse fullcircle scaled 1w shifted (0.5w,0.5h)}
\end{fmfgraph}
}}
\, .
\end{eqnarray}
Expanding again the exponential in a power series, we obtain
\begin{eqnarray}
\label{NANA}
Z &= & \left\{ 1 + \frac{-1}{2} \, \int_{1234} V_{1234}^{({\rm int})} \, 
\frac{\delta^2}{\delta G^{-1}_{12} \delta G^{-1}_{34}} + 
\frac{1}{2} \left(  \frac{-1}{2} \right)^2 \int_{12345678}
V_{1234}^{({\rm int})} V_{5678}^{({\rm int})}  
\frac{\delta^4}{\delta G^{-1}_{12} \delta G^{-1}_{34}
\delta G^{-1}_{56} \delta G^{-1}_{78}} + \ldots \right\}
e^{\Omega^{(0)}} \, .
\end{eqnarray}
Note the two advantages of this expansion over the conventional one (\ref{EJ}) in terms of currents coupled
linearly to the fields. First, it contains only half as many functional derivatives. Second, in case of fermions it does not contain
derivatives with respect to Grassmann fields.
Taking into account (\ref{DR1}), (\ref{ACT}), and (\ref{LOG}), we obtain
\begin{eqnarray}
\label{PROP}
\frac{\delta}{\delta G^{-1}_{12}} \, e^{\Omega^{(0)}} &=& \mp \, G_{21} \, e^{\Omega^{(0)}} \, , \\ 
\frac{\delta^2}{\delta G^{-1}_{12} \delta G^{-1}_{34}} \, e^{\Omega^{(0)}} &=&
\left( G_{21} G_{43} \pm G_{23} G_{41} \right) \, e^{\Omega^{(0)}} \, ,
\label{PROPB}
\end{eqnarray}
such that the partition function $Z$ becomes
\begin{eqnarray}
Z & = & \left\{ 1 - \frac{1}{2} \, \int_{1234} V_{1234}^{({\rm int})} \, \left( G_{21} G_{43} \pm G_{23} G_{41} \right) 
\, + \, \ldots \right\} \, e^{\Omega^{(0)}} \, . 
\end{eqnarray}
According to the Feynman rules (\ref{PRO}) and (\ref{VE}), 
this is represented by the diagrams
\begin{eqnarray}
\label{W1}
Z = \left\{ 1 +  \frac{1}{2} \,
\setlength{\unitlength}{1mm}
\parbox{19mm}{\centerline{
\begin{fmfgraph}(15,7)
\setval
\fmfforce{0.33w,0.5h}{v1}
\fmfforce{0.66w,0.5h}{v2}
\fmf{dashes}{v1,v2}
\fmfi{fermion}{reverse fullcircle scaled 0.33w shifted (0.165w,0.5h)}
\fmfi{fermion}{fullcircle rotated 180 scaled 0.33w shifted (0.825w,0.5h)}
\fmfdot{v1,v2}
\end{fmfgraph}
}}
\, \pm\, \frac{1}{2} \, 
\setlength{\unitlength}{1mm}
\parbox{10mm}{\centerline{
\begin{fmfgraph}(7,5)
\setval
\fmfleft{v1}
\fmfright{v2}
\fmf{dashes}{v1,v2}
\fmf{fermion,left=0.7}{v2,v1}
\fmf{fermion,left=0.7}{v1,v2}
\fmfdot{v1,v2}
\end{fmfgraph}
}} 
\, + \, \ldots \right\} \, 
\exp \left\{
\mp
\setlength{\unitlength}{1mm}
\parbox{8mm}{\centerline{
\begin{fmfgraph}(5,5)
\setval
\fmfi{fermion}{reverse fullcircle scaled 1w shifted (0.5w,0.5h)}
\end{fmfgraph}
}}
\right\} \hspace*{2mm} .
\end{eqnarray}
Note that each closed loop causes a factor $\pm 1$. 
\section{Graphical Recursion Relation for Connected Vacuum Diagrams} \label{graph}
In this section, we derive a functional differential equation for the grand-canonical potential 
$\Omega$ whose solution leads to a graphical
recursion relation for all connected vacuum diagrams.
\subsection{Functional Differential Equation}
The functional differential equation for the grand-canonical potential 
$\Omega$ is derived from the following functional integral identity
\begin{equation}
\label{FI} 
\int {\cal D} \psi^* \int {\cal D} \psi \,  \frac{\delta}{\delta \psi_1^*}\left( \psi_2^* \,  
e^{-{\cal A}[\psi^*,\psi]} \right)=0
\end{equation}
with the action (\ref{AF}). This identity is the functional 
generalization of the trivial integral identity 
$\int_{-\infty}^{+\infty} dx\,f'(x)=0$ for
functions $f(x)$ which vanish at infinity. Nontrivial consequences of
Eq.~(\ref{FI}) are obtained by
performing the functional derivative in the integrand which yields
\begin{eqnarray}
\label{FI2}
\int {\cal D} \psi^* \int {\cal D} \psi
\left( \delta_{12} \mp
\int_3 G^{-1}_{13} \psi^*_2 \psi_3 \mp \int_{345} V_{1345}^{({\rm int})} \psi_2^* \psi_3 \psi^*_4 \psi_5 \right)
e^{-{\cal A}[\psi^*,\psi]} =0 \,.
\end{eqnarray}
Substituting the field product $\psi_2^* \psi_3$ by a functional 
derivative with respect to the kernel $G^{-1}_{23}$ according to 
\begin{eqnarray}
\psi_2^* \psi_3 \,\, e^{-{\cal A}[\psi^*,\psi]} = - \frac{\delta}{\delta G_{23}^{-1}} \,\,e^{-{\cal A}[\psi^*,\psi]}
\,, 
\end{eqnarray}
this equation can be expressed in terms of the partition function (\ref{PF}):
\begin{eqnarray}
\label{FUNC1}
Z \, \delta_{12} \pm \int_3 G^{-1}_{13} \, \frac{\delta Z}{\delta G^{-1}_{23}}
= \pm \int_{345} V_{1345}^{({\rm int})} \, \frac{\delta^2 Z}{\delta G^{-1}_{23}\delta G^{-1}_{45}}\, .
\end{eqnarray}
Going over from the partition function $Z$ to the grand-canonical potential 
\begin{eqnarray}
\Omega = \ln Z \, ,
\end{eqnarray}
the linear functional differential equation (\ref{FUNC1}) turns into a nonlinear one:
\begin{eqnarray}
\label{FUNC2}
\delta_{12} \pm \int_3 G^{-1}_{13} \, \frac{\delta \Omega}{\delta G^{-1}_{23}}
= \pm \int_{345} V_{1345}^{({\rm int})} \, \left( \frac{\delta^2 \Omega}{\delta G^{-1}_{23}\delta G^{-1}_{45}}
+ \frac{\delta \Omega}{\delta G^{-1}_{23}} \, \frac{\delta \Omega}{\delta G^{-1}_{45}}
\right) \, .
\end{eqnarray}
If the interaction potential $V^{({\rm int})}$ vanishes, this is immediately 
solved by (\ref{FPF}) due to (\ref{PROP}). For a non-vanishing interaction potential $V^{({\rm int})}$, the
right-hand side in (\ref{FUNC2}) produces corrections to (\ref{FPF})
which we shall denote with $\Omega^{({\rm int})}$. Thus
the grand-canonical potential $\Omega$ decomposes according to
\begin{eqnarray}
\label{DEC}
\Omega = \Omega^{(0)} + \Omega^{({\rm int})} \, .
\end{eqnarray}
Inserting this into (\ref{FUNC2}) and taking into account (\ref{PROP}) as well as (\ref{PROPB}),
we obtain the following functional differential equation for the interaction
part of the grand-canonical potential $\Omega^{({\rm int})}$:
\begin{eqnarray}
\label{FUNC3}
\pm \int_3 G^{-1}_{13} \, \frac{\delta \Omega^{({\rm int})}}{\delta G^{-1}_{23}}
&=& \int_{345}  V_{1345}^{({\rm int})} G_{34} G_{52} \pm \int_{345} V_{1345}^{({\rm int})} G_{32} G_{54}
- \int_{345} V_{1345}^{({\rm int})} G_{32} \frac{\delta \Omega^{({\rm int})}}{\delta G^{-1}_{45}} \nonumber \\
&&- \int_{345} V_{1345}^{({\rm int})} G_{54} \frac{\delta \Omega^{({\rm int})}}{\delta G^{-1}_{23}}
\pm \int_{345} V_{1345}^{({\rm int})} \, \left( \, \frac{\delta^2 \Omega^{({\rm int})}}{\delta G^{-1}_{23}\delta G^{-1}_{45}}
\pm \frac{\delta \Omega^{({\rm int})}}{\delta G^{-1}_{23}} \, \frac{\delta \Omega^{({\rm int})}}{\delta G^{-1}_{45}}
\right) \, .
\end{eqnarray}
Setting $1=2$ and performing the integration over $1$, we obtain because of the symmetry (\ref{SYM})
of the interaction potential $V^{({\rm int})}$:
\begin{eqnarray}
\label{FUNC4}
- \int_{12} G^{-1}_{12} \, \frac{\delta \Omega^{({\rm int})}}{\delta G^{-1}_{12}}
&= & - \int_{1234} V^{({\rm int})}_{1234} G_{21} G_{43} \mp \int_{1234} V^{({\rm int})}_{1234} G_{41} G_{23}
\pm 2 \int_{1234} V_{1234}^{({\rm int})} G_{21} \frac{\delta \Omega^{({\rm int})}}{\delta G^{-1}_{34}} \nonumber \\
&& - \int_{1234} V_{1234}^{({\rm int})} \, \frac{\delta^2 \Omega^{({\rm int})}}{\delta G^{-1}_{12}\delta G^{-1}_{34}}
- \int_{1234} V_{1234}^{({\rm int})} \, 
\frac{\delta \Omega^{({\rm int})}}{\delta G^{-1}_{12}} \, \frac{\delta \Omega^{({\rm int})}}{\delta G^{-1}_{34}}
\, .
\end{eqnarray}
Equation (\ref{FUNC4}) contains 
functional derivatives with 
respect to the kernel $G^{-1}$ which are equivalent
to cutting lines in the vacuum diagrams.
For practical purposes, however, it will be 
more convenient to work with
derivatives with respect to the 
correlation functions $G$ which remove lines.
With the help of the functional chain rule, the first and second 
derivatives with respect to the kernel $G^{-1}$ are rewritten as (\ref{CHAIN1}) and
\begin{eqnarray}
\label{CHAIN2}
\hspace*{-1cm} \frac{\delta^2}{\delta G^{-1}_{12} \delta G^{-1}_{34}} =  \int_{5678}
G_{71} G_{28} G_{53} G_{46} \frac{\delta^2}{\delta G_{78} \delta G_{56}} 
+ \int_{56} \left( G_{51} G_{23} G_{46} +
G_{41} G_{26} G_{53} \right) \frac{\delta}{\delta G_{56}} \, ,
\end{eqnarray}
respectively. Taking into account again the symmetry (\ref{SYM})
of the interaction potential $V^{({\rm int})}$, 
the functional differential equation (\ref{FUNC4}) for  
$\Omega^{({\rm int})}$ takes the final form
\begin{eqnarray}
\label{FUNC5}
\hspace*{-0.5cm} \int_{12} G_{12} \, \frac{\delta \Omega^{({\rm int})}}{\delta G_{12}}
&=&  - \int_{1234} V^{({\rm int})}_{1234} G_{21} G_{43} \mp \int_{1234} V^{({\rm int})}_{1234} G_{23} G_{41}
\nonumber \\ && 
- 2 \int_{123456} V_{1234}^{({\rm int})} G_{51} G_{23} G_{46} \frac{\delta \Omega^{({\rm int})}}{\delta G_{56}} 
\mp 2 \int_{123456} V_{1234}^{({\rm int})} G_{21} G_{53} G_{46} \frac{\delta \Omega^{({\rm int})}}{\delta G_{56}}
\nonumber \\ &&   
- \int_{12345678} V_{1234}^{({\rm int})} G_{71} G_{28} G_{53} G_{46} \frac{\delta^2 \Omega^{({\rm int})}}{\delta G_{78}\delta G_{56}}
- \int_{12345678} V_{1234}^{({\rm int})} G_{71} G_{28} G_{53} G_{46} \frac{\delta \Omega^{({\rm int})}}{\delta G_{78}} 
\frac{\delta \Omega^{({\rm int})}}{\delta G_{56}} \, .
\end{eqnarray}
Supplementing the Feynman rules (\ref{PRO}) and (\ref{VE}) with a graphical representation of the interaction part of the
grand-canonical potential
\begin{eqnarray}
\Omega^{({\rm int})} \,\, = \,\,
\setlength{\unitlength}{1mm}
\parbox{10mm}{\centerline{
  \begin{fmfgraph*}(8,6)
  \setval
  \fmfforce{0w,1/2h}{v1}
  \fmfforce{1w,1/2h}{v2}
  \fmfforce{1/2w,1h}{v3}
  \fmfforce{1/2w,0h}{v4}
  \fmfforce{1/2w,1/2h}{v5}
  \fmf{plain,left=0.4}{v1,v3,v2,v4,v1}
  \end{fmfgraph*} } }  \,\,\, ,
\end{eqnarray}
the functional differential equation (\ref{FUNC5}) can be depicted graphically as follows:
\begin{eqnarray}
\setlength{\unitlength}{1mm}
\parbox{5.5mm}{\begin{center}
\begin{fmfgraph*}(2.5,5)
\setval
\fmfstraight
\fmfforce{1w,0h}{v1}
\fmfforce{1w,1h}{v2}
\fmf{fermion,left=1}{v1,v2}
\fmfv{decor.size=0, label=${\scriptstyle 2}$, l.dist=1mm, l.angle=0}{v1}
\fmfv{decor.size=0, label=${\scriptstyle 1}$, l.dist=1mm, l.angle=0}{v2}
\end{fmfgraph*}
\end{center}}
\hspace*{0.2cm} \dphi{
\parbox{10mm}{\centerline{
  \begin{fmfgraph*}(8,6)
  \setval
  \fmfforce{0w,1/2h}{v1}
  \fmfforce{1w,1/2h}{v2}
  \fmfforce{1/2w,1h}{v3}
  \fmfforce{1/2w,0h}{v4}
  \fmfforce{1/2w,1/2h}{v5}
  \fmf{plain,left=0.4}{v1,v3,v2,v4,v1}
  \end{fmfgraph*} } } 
}{1}{2} 
& = &
\setlength{\unitlength}{1mm}
\parbox{17mm}{\centerline{
\begin{fmfgraph}(15,7)
\setval
\fmfforce{0.33w,0.5h}{v1}
\fmfforce{0.66w,0.5h}{v2}
\fmf{dashes}{v1,v2}
\fmfi{fermion}{reverse fullcircle scaled 0.33w shifted (0.165w,0.5h)}
\fmfi{fermion}{fullcircle rotated 180 scaled 0.33w shifted (0.825w,0.5h)}
\fmfdot{v1,v2}
\end{fmfgraph}
}}
\, \pm \, 
\setlength{\unitlength}{1mm}
\parbox{10mm}{\centerline{
\begin{fmfgraph}(7,5)
\setval
\fmfleft{v1}
\fmfright{v2}
\fmf{dashes}{v1,v2}
\fmf{fermion,left=0.7}{v2,v1}
\fmf{fermion,left=0.7}{v1,v2}
\fmfdot{v1,v2}
\end{fmfgraph}
}} 
\, + \, 2 \, 
\setlength{\unitlength}{1mm}
\parbox{13mm}{\begin{center}
\begin{fmfgraph*}(8,6)
\setval
\fmfstraight
\fmfforce{0.3w,1h}{i1}
\fmfforce{0.3w,0h}{i2}
\fmfforce{1w,1h}{o1}
\fmfforce{1w,0h}{o2}
\fmf{fermion}{i2,i1}
\fmf{dashes,right=0.7}{i1,i2}
\fmf{fermion}{i1,o1}
\fmf{fermion}{o2,i2}
\fmfdot{i1,i2}
\fmfv{decor.size=0, label=${\scriptstyle 1}$, l.dist=1mm, l.angle=0}{o1}
\fmfv{decor.size=0, label=${\scriptstyle 2}$, l.dist=1mm, l.angle=0}{o2}
\end{fmfgraph*}
\end{center}}
\hspace*{1mm} \dphi{
\setlength{\unitlength}{1mm}
\parbox{10mm}{\centerline{
  \begin{fmfgraph*}(8,6)
  \setval
  \fmfforce{0w,1/2h}{v1}
  \fmfforce{1w,1/2h}{v2}
  \fmfforce{1/2w,1h}{v3}
  \fmfforce{1/2w,0h}{v4}
  \fmfforce{1/2w,1/2h}{v5}
  \fmf{plain,left=0.4}{v1,v3,v2,v4,v1}
  \end{fmfgraph*} } } 
}{1}{2}
\, \pm \, 2 \, 
\setlength{\unitlength}{1mm}
\parbox{17mm}{\begin{center}
\begin{fmfgraph*}(13,6)
\setval
\fmfstraight
\fmfforce{5/13w,0.5h}{v1}
\fmfforce{10/13w,0.5h}{v2}
\fmfforce{2.5/13w,0.5/6h}{v3}
\fmfforce{2.5/13w,5.5/6h}{v4}
\fmfforce{1w,1h}{o1}
\fmfforce{1w,0h}{o2}
\fmf{dashes}{v1,v2}
\fmf{fermion,left=0.1}{v2,o1}
\fmf{fermion,left=0.1}{o2,v2}
\fmf{fermion,left=1}{v3,v4}
\fmf{plain,left=1}{v4,v3}
\fmfdot{v1,v2}
\fmfv{decor.size=0, label=${\scriptstyle 1}$, l.dist=1mm, l.angle=0}{o1}
\fmfv{decor.size=0, label=${\scriptstyle 2}$, l.dist=1mm, l.angle=0}{o2}
\end{fmfgraph*}
\end{center}}
\hspace*{4mm} \dphi{
\setlength{\unitlength}{1mm}
\parbox{10mm}{\centerline{
  \begin{fmfgraph*}(8,6)
  \setval
  \fmfforce{0w,1/2h}{v1}
  \fmfforce{1w,1/2h}{v2}
  \fmfforce{1/2w,1h}{v3}
  \fmfforce{1/2w,0h}{v4}
  \fmfforce{1/2w,1/2h}{v5}
  \fmf{plain,left=0.4}{v1,v3,v2,v4,v1}
  \end{fmfgraph*} } } 
}{1}{2} \nonumber \\
&& + \,  
\setlength{\unitlength}{1mm}
\parbox{14mm}{\begin{center}
\begin{fmfgraph*}(8,9)
\setval
\fmfstraight
\fmfforce{0w,0.835h}{i1}
\fmfforce{0w,0.165h}{i2}
\fmfforce{1w,1h}{o1}
\fmfforce{1w,0.66h}{o2}
\fmfforce{1w,0.33h}{o3}
\fmfforce{1w,0h}{o4}
\fmf{dashes}{i1,i2}
\fmf{fermion,left=0.1}{i1,o1}
\fmf{fermion,left=0.1}{o2,i1}
\fmf{fermion,left=0.1}{i2,o3}
\fmf{fermion,left=0.1}{o4,i2}
\fmfdot{i1,i2}
\fmfv{decor.size=0, label=${\scriptstyle 1}$, l.dist=1mm, l.angle=0}{o1}
\fmfv{decor.size=0, label=${\scriptstyle 2}$, l.dist=1mm, l.angle=0}{o2}
\fmfv{decor.size=0, label=${\scriptstyle 3}$, l.dist=1mm, l.angle=0}{o3}
\fmfv{decor.size=0, label=${\scriptstyle 4}$, l.dist=1mm, l.angle=0}{o4}
\end{fmfgraph*}
\end{center}}
\hspace*{1mm} \ddphi{
\setlength{\unitlength}{1mm}
\parbox{10mm}{\centerline{
  \begin{fmfgraph*}(8,6)
  \setval
  \fmfforce{0w,1/2h}{v1}
  \fmfforce{1w,1/2h}{v2}
  \fmfforce{1/2w,1h}{v3}
  \fmfforce{1/2w,0h}{v4}
  \fmfforce{1/2w,1/2h}{v5}
  \fmf{plain,left=0.4}{v1,v3,v2,v4,v1}
  \end{fmfgraph*} } } 
}
\, + \, \dphi{
\setlength{\unitlength}{1mm}
\parbox{10mm}{\centerline{
  \begin{fmfgraph*}(8,6)
  \setval
  \fmfforce{0w,1/2h}{v1}
  \fmfforce{1w,1/2h}{v2}
  \fmfforce{1/2w,1h}{v3}
  \fmfforce{1/2w,0h}{v4}
  \fmfforce{1/2w,1/2h}{v5}
  \fmf{plain,left=0.4}{v1,v3,v2,v4,v1}
  \end{fmfgraph*} } } 
}{1}{2} \hspace*{3mm}
\parbox{17mm}{\begin{center}
\begin{fmfgraph*}(15,6)
\setval
\fmfstraight
\fmfforce{0w,1h}{i1}
\fmfforce{0w,0h}{i2}
\fmfforce{1/3w,1/2h}{v1}
\fmfforce{2/3w,1/2h}{v2}
\fmfforce{1w,1h}{o1}
\fmfforce{1w,0h}{o2}
\fmf{fermion,right=0.1}{v1,i1}
\fmf{fermion,right=0.1}{i2,v1}
\fmf{dashes}{v1,v2}
\fmf{fermion,left=0.1}{v2,o1}
\fmf{fermion,left=0.1}{o2,v2}
\fmfdot{v1,v2}
\fmfv{decor.size=0, label=${\scriptstyle 1}$, l.dist=1mm, l.angle=-180}{i1}
\fmfv{decor.size=0, label=${\scriptstyle 2}$, l.dist=1mm, l.angle=-180}{i2}
\fmfv{decor.size=0, label=${\scriptstyle 3}$, l.dist=1mm, l.angle=0}{o1}
\fmfv{decor.size=0, label=${\scriptstyle 4}$, l.dist=1mm, l.angle=0}{o2}
\end{fmfgraph*}
\end{center}}
\hspace*{3mm} \dphi{
\setlength{\unitlength}{1mm}
\parbox{10mm}{\centerline{
  \begin{fmfgraph*}(8,6)
  \setval
  \fmfforce{0w,1/2h}{v1}
  \fmfforce{1w,1/2h}{v2}
  \fmfforce{1/2w,1h}{v3}
  \fmfforce{1/2w,0h}{v4}
  \fmfforce{1/2w,1/2h}{v5}
  \fmf{plain,left=0.4}{v1,v3,v2,v4,v1}
  \end{fmfgraph*} } } 
}{3}{4} \, . 
\label{FUNC6}
\end{eqnarray}
\subsection{Recursion Relation}
We now convert the functional differential equation (\ref{FUNC6}) into a recursion
relation by expanding the interaction part of the grand-canonical potential 
into a power series of the interaction potential $V^{({\rm int})}$:
\begin{eqnarray}
\label{EXP}
\setlength{\unitlength}{1mm}
\parbox{10mm}{\centerline{
  \begin{fmfgraph*}(8,6)
  \setval
  \fmfforce{0w,1/2h}{v1}
  \fmfforce{1w,1/2h}{v2}
  \fmfforce{1/2w,1h}{v3}
  \fmfforce{1/2w,0h}{v4}
  \fmfforce{1/2w,1/2h}{v5}
  \fmf{plain,left=0.4}{v1,v3,v2,v4,v1}
  \end{fmfgraph*} } } 
= \sum_{p = 1}^{\infty} 
\parbox{10mm}{\centerline{
  \begin{fmfgraph*}(8,6)
  \setval
  \fmfforce{0w,1/2h}{v1}
  \fmfforce{1w,1/2h}{v2}
  \fmfforce{1/2w,1h}{v3}
  \fmfforce{1/2w,0h}{v4}
  \fmfforce{1/2w,1/2h}{v5}
  \fmf{plain,left=0.4}{v1,v3,v2,v4,v1}
  \fmfv{decor.size=0, label=${\scriptstyle p}$, l.dist=0mm, l.angle=0}{v5}
  \end{fmfgraph*} } }  
\, .
\end{eqnarray}
The vacuum diagrams which contain $p$ times the interaction potential $V^{({\rm int})}$ 
have the property that they satisfy the eigenvalue problem
\begin{eqnarray}
\label{NUMB}
\setlength{\unitlength}{1mm}
\parbox{5.5mm}{\begin{center}
\begin{fmfgraph*}(2.5,5)
\setval
\fmfstraight
\fmfforce{1w,0h}{v1}
\fmfforce{1w,1h}{v2}
\fmf{fermion,left=1}{v1,v2}
\fmfv{decor.size=0, label=${\scriptstyle 2}$, l.dist=1mm, l.angle=0}{v1}
\fmfv{decor.size=0, label=${\scriptstyle 1}$, l.dist=1mm, l.angle=0}{v2}
\end{fmfgraph*}
\end{center}}
\hspace*{0.2cm} \dphi{
\parbox{10mm}{\centerline{
  \begin{fmfgraph*}(8,6)
  \setval
  \fmfforce{0w,1/2h}{v1}
  \fmfforce{1w,1/2h}{v2}
  \fmfforce{1/2w,1h}{v3}
  \fmfforce{1/2w,0h}{v4}
  \fmfforce{1/2w,1/2h}{v5}
  \fmf{plain,left=0.4}{v1,v3,v2,v4,v1}
  \fmfv{decor.size=0, label=${\scriptstyle p}$, l.dist=0mm, l.angle=0}{v5}
  \end{fmfgraph*} } }  
}{1}{2} = 2 p \, 
\parbox{10mm}{\centerline{
  \begin{fmfgraph*}(8,6)
  \setval
  \fmfforce{0w,1/2h}{v1}
  \fmfforce{1w,1/2h}{v2}
  \fmfforce{1/2w,1h}{v3}
  \fmfforce{1/2w,0h}{v4}
  \fmfforce{1/2w,1/2h}{v5}
  \fmf{plain,left=0.4}{v1,v3,v2,v4,v1}
  \fmfv{decor.size=0, label=${\scriptstyle p}$, l.dist=0mm, l.angle=0}{v5}
  \end{fmfgraph*} } }  
 \, .
\end{eqnarray}
As the operator on the left-hand side removes a line from the Feynman diagram which is later on restored,
it counts the number $2 p$ of lines in a vacuum diagram. Inserting the expansion (\ref{EXP}) into the functional
differential equation (\ref{FUNC6}), we obtain for $p\ge 1$ the recursion relation
\begin{eqnarray}
\setlength{\unitlength}{1mm}
\parbox{10mm}{\centerline{
  \begin{fmfgraph*}(8,6)
  \setval
  \fmfforce{0w,1/2h}{v1}
  \fmfforce{1w,1/2h}{v2}
  \fmfforce{1/2w,1h}{v3}
  \fmfforce{1/2w,0h}{v4}
  \fmfforce{1/2w,1/2h}{v5}
  \fmf{plain,left=0.4}{v1,v3,v2,v4,v1}
  \fmfv{decor.size=0, label=${\scriptstyle p+1}$, l.dist=0mm, l.angle=0}{v5}
  \end{fmfgraph*} } }   
& = & \frac{1}{2(p+1)} \left(
\, 2 \, 
\setlength{\unitlength}{1mm}
\parbox{13mm}{\begin{center}
\begin{fmfgraph*}(8,6)
\setval
\fmfstraight
\fmfforce{0.3w,1h}{i1}
\fmfforce{0.3w,0h}{i2}
\fmfforce{1w,1h}{o1}
\fmfforce{1w,0h}{o2}
\fmf{fermion}{i2,i1}
\fmf{dashes,right=0.7}{i1,i2}
\fmf{fermion}{i1,o1}
\fmf{fermion}{o2,i2}
\fmfdot{i1,i2}
\fmfv{decor.size=0, label=${\scriptstyle 1}$, l.dist=1mm, l.angle=0}{o1}
\fmfv{decor.size=0, label=${\scriptstyle 2}$, l.dist=1mm, l.angle=0}{o2}
\end{fmfgraph*}
\end{center}}
\hspace*{1mm} \dphi{
\parbox{10mm}{\centerline{
  \begin{fmfgraph*}(8,6)
  \setval
  \fmfforce{0w,1/2h}{v1}
  \fmfforce{1w,1/2h}{v2}
  \fmfforce{1/2w,1h}{v3}
  \fmfforce{1/2w,0h}{v4}
  \fmfforce{1/2w,1/2h}{v5}
  \fmf{plain,left=0.4}{v1,v3,v2,v4,v1}
  \fmfv{decor.size=0, label=${\scriptstyle p}$, l.dist=0mm, l.angle=0}{v5}
  \end{fmfgraph*} } }  
}{1}{2}
\, \pm \, 2 \, 
\setlength{\unitlength}{1mm}
\parbox{17mm}{\begin{center}
\begin{fmfgraph*}(13,6)
\setval
\fmfstraight
\fmfforce{5/13w,0.5h}{v1}
\fmfforce{10/13w,0.5h}{v2}
\fmfforce{2.5/13w,0.5/6h}{v3}
\fmfforce{2.5/13w,5.5/6h}{v4}
\fmfforce{1w,1h}{o1}
\fmfforce{1w,0h}{o2}
\fmf{dashes}{v1,v2}
\fmf{fermion,left=0.1}{v2,o1}
\fmf{fermion,left=0.1}{o2,v2}
\fmf{fermion,left=1}{v3,v4}
\fmf{plain,left=1}{v4,v3}
\fmfdot{v1,v2}
\fmfv{decor.size=0, label=${\scriptstyle 1}$, l.dist=1mm, l.angle=0}{o1}
\fmfv{decor.size=0, label=${\scriptstyle 2}$, l.dist=1mm, l.angle=0}{o2}
\end{fmfgraph*}
\end{center}}
\hspace*{4mm} \dphi{
\parbox{10mm}{\centerline{
  \begin{fmfgraph*}(8,6)
  \setval
  \fmfforce{0w,1/2h}{v1}
  \fmfforce{1w,1/2h}{v2}
  \fmfforce{1/2w,1h}{v3}
  \fmfforce{1/2w,0h}{v4}
  \fmfforce{1/2w,1/2h}{v5}
  \fmf{plain,left=0.4}{v1,v3,v2,v4,v1}
  \fmfv{decor.size=0, label=${\scriptstyle p}$, l.dist=0mm, l.angle=0}{v5}
  \end{fmfgraph*} } } 
}{1}{2}
\right. \nonumber \\
&& \left. + \,  
\setlength{\unitlength}{1mm}
\parbox{14mm}{\begin{center}
\begin{fmfgraph*}(8,9)
\setval
\fmfstraight
\fmfforce{0w,0.835h}{i1}
\fmfforce{0w,0.165h}{i2}
\fmfforce{1w,1h}{o1}
\fmfforce{1w,0.66h}{o2}
\fmfforce{1w,0.33h}{o3}
\fmfforce{1w,0h}{o4}
\fmf{dashes}{i1,i2}
\fmf{fermion,left=0.1}{i1,o1}
\fmf{fermion,left=0.1}{o2,i1}
\fmf{fermion,left=0.1}{i2,o3}
\fmf{fermion,left=0.1}{o4,i2}
\fmfdot{i1,i2}
\fmfv{decor.size=0, label=${\scriptstyle 1}$, l.dist=1mm, l.angle=0}{o1}
\fmfv{decor.size=0, label=${\scriptstyle 2}$, l.dist=1mm, l.angle=0}{o2}
\fmfv{decor.size=0, label=${\scriptstyle 3}$, l.dist=1mm, l.angle=0}{o3}
\fmfv{decor.size=0, label=${\scriptstyle 4}$, l.dist=1mm, l.angle=0}{o4}
\end{fmfgraph*}
\end{center}}
\hspace*{1mm} \ddphi{
\parbox{10mm}{\centerline{
  \begin{fmfgraph*}(8,6)
  \setval
  \fmfforce{0w,1/2h}{v1}
  \fmfforce{1w,1/2h}{v2}
  \fmfforce{1/2w,1h}{v3}
  \fmfforce{1/2w,0h}{v4}
  \fmfforce{1/2w,1/2h}{v5}
  \fmf{plain,left=0.4}{v1,v3,v2,v4,v1}
  \fmfv{decor.size=0, label=${\scriptstyle p}$, l.dist=0mm, l.angle=0}{v5}
  \end{fmfgraph*} } } 
}
\, + \sum_{q=1}^{p-1} \, \dphi{
\parbox{10mm}{\centerline{
  \begin{fmfgraph*}(8,6)
  \setval
  \fmfforce{0w,1/2h}{v1}
  \fmfforce{1w,1/2h}{v2}
  \fmfforce{1/2w,1h}{v3}
  \fmfforce{1/2w,0h}{v4}
  \fmfforce{1/2w,1/2h}{v5}
  \fmf{plain,left=0.4}{v1,v3,v2,v4,v1}
  \fmfv{decor.size=0, label=${\scriptstyle q}$, l.dist=0mm, l.angle=0}{v5}
  \end{fmfgraph*} } } 
}{1}{2} \hspace*{3mm}
\parbox{17mm}{\begin{center}
\begin{fmfgraph*}(15,6)
\setval
\fmfstraight
\fmfforce{0w,1h}{i1}
\fmfforce{0w,0h}{i2}
\fmfforce{1/3w,1/2h}{v1}
\fmfforce{2/3w,1/2h}{v2}
\fmfforce{1w,1h}{o1}
\fmfforce{1w,0h}{o2}
\fmf{fermion,right=0.1}{v1,i1}
\fmf{fermion,right=0.1}{i2,v1}
\fmf{dashes}{v1,v2}
\fmf{fermion,left=0.1}{v2,o1}
\fmf{fermion,left=0.1}{o2,v2}
\fmfdot{v1,v2}
\fmfv{decor.size=0, label=${\scriptstyle 1}$, l.dist=1mm, l.angle=-180}{i1}
\fmfv{decor.size=0, label=${\scriptstyle 2}$, l.dist=1mm, l.angle=-180}{i2}
\fmfv{decor.size=0, label=${\scriptstyle 3}$, l.dist=1mm, l.angle=0}{o1}
\fmfv{decor.size=0, label=${\scriptstyle 4}$, l.dist=1mm, l.angle=0}{o2}
\end{fmfgraph*}
\end{center}}
\hspace*{3mm} \dphi{
\parbox{10mm}{\centerline{
  \begin{fmfgraph*}(8,6)
  \setval
  \fmfforce{0w,1/2h}{v1}
  \fmfforce{1w,1/2h}{v2}
  \fmfforce{1/2w,1h}{v3}
  \fmfforce{1/2w,0h}{v4}
  \fmfforce{1/2w,1/2h}{v5}
  \fmf{plain,left=0.4}{v1,v3,v2,v4,v1}
  \fmfv{decor.size=0, label=${\scriptstyle p-q}$, l.dist=0mm, l.angle=0}{v5}
  \end{fmfgraph*} } } 
}{3}{4}  \right) 
\label{FUNC7}
\end{eqnarray}
and the initial diagrams 
\begin{eqnarray}
\setlength{\unitlength}{1mm}
\parbox{10mm}{\centerline{
  \begin{fmfgraph*}(8,6)
  \setval
  \fmfforce{0w,1/2h}{v1}
  \fmfforce{1w,1/2h}{v2}
  \fmfforce{1/2w,1h}{v3}
  \fmfforce{1/2w,0h}{v4}
  \fmfforce{1/2w,1/2h}{v5}
  \fmf{plain,left=0.4}{v1,v3,v2,v4,v1}
  \fmfv{decor.size=0, label=${\scriptstyle 1}$, l.dist=0mm, l.angle=0}{v5}
  \end{fmfgraph*} } }  
=  \frac{1}{2} \,
\setlength{\unitlength}{1mm}
\parbox{19mm}{\centerline{
\begin{fmfgraph}(15,7)
\setval
\fmfforce{0.33w,0.5h}{v1}
\fmfforce{0.66w,0.5h}{v2}
\fmf{dashes}{v1,v2}
\fmfi{fermion}{reverse fullcircle scaled 0.33w shifted (0.165w,0.5h)}
\fmfi{fermion}{fullcircle rotated 180 scaled 0.33w shifted (0.825w,0.5h)}
\fmfdot{v1,v2}
\end{fmfgraph}
}}
\, \pm\, \frac{1}{2} \, 
\setlength{\unitlength}{1mm}
\parbox{10mm}{\centerline{
\begin{fmfgraph}(7,5)
\setval
\fmfleft{v1}
\fmfright{v2}
\fmf{dashes}{v1,v2}
\fmf{fermion,left=0.7}{v2,v1}
\fmf{fermion,left=0.7}{v1,v2}
\fmfdot{v1,v2}
\end{fmfgraph}
}} 
\hspace*{2mm} .
\label{BEG}
\end{eqnarray}
The right-hand side of (\ref{FUNC7}) contains four graphical operations. The first three are linear
and involve one or two line amputations of the previous perturbative
order. The fourth operation is nonlinear and mixes two different
line amputations of lower orders.
To demonstrate the working of this recursion formula, we construct 
the connected vacuum diagrams in second and third order. 
We start with the amputation of one or two lines in the first order (\ref{BEG}):
\begin{eqnarray}
\label{O2}
\dphi{
\setlength{\unitlength}{1mm}
\parbox{10mm}{\centerline{
  \begin{fmfgraph*}(8,6)
  \setval
  \fmfforce{0w,1/2h}{v1}
  \fmfforce{1w,1/2h}{v2}
  \fmfforce{1/2w,1h}{v3}
  \fmfforce{1/2w,0h}{v4}
  \fmfforce{1/2w,1/2h}{v5}
  \fmf{plain,left=0.4}{v1,v3,v2,v4,v1}
  \fmfv{decor.size=0, label=${\scriptstyle 1}$, l.dist=0mm, l.angle=0}{v5}
  \end{fmfgraph*} } } 
}{1}{2}  =  
\setlength{\unitlength}{1mm}
\parbox{17mm}{\begin{center}
\begin{fmfgraph*}(12,6)
\setval
\fmfstraight
\fmfforce{2/12w,0.5h}{v1}
\fmfforce{7/12w,0.5h}{v2}
\fmfforce{9.5/12w,0.5/6h}{v3}
\fmfforce{9.5/12w,5.5/6h}{v4}
\fmfforce{0w,5/6h}{i1}
\fmfforce{0w,1/6h}{i2}
\fmf{dashes}{v1,v2}
\fmf{fermion}{i1,v1}
\fmf{fermion}{v1,i2}
\fmf{fermion,left=1}{v4,v3}
\fmf{plain,left=1}{v3,v4}
\fmfdot{v1,v2}
\fmfv{decor.size=0, label=${\scriptstyle 1}$, l.dist=1mm, l.angle=-180}{i1}
\fmfv{decor.size=0, label=${\scriptstyle 2}$, l.dist=1mm, l.angle=-180}{i2}
\end{fmfgraph*}
\end{center}}
\;\pm\quad
\setlength{\unitlength}{1mm}
\parbox{5mm}{\begin{center}
\begin{fmfgraph*}(3,6)
\setval
\fmfstraight
\fmfforce{0.7w,1h}{o1}
\fmfforce{0.7w,0h}{o2}
\fmfforce{-1/3w,1h}{i1}
\fmfforce{-1/3w,0h}{i2}
\fmf{fermion}{o1,o2}
\fmf{dashes,left=0.7}{o1,o2}
\fmf{fermion}{i1,o1}
\fmf{fermion}{o2,i2}
\fmfdot{o1,o2}
\fmfv{decor.size=0, label=${\scriptstyle 1}$, l.dist=1mm, l.angle=-180}{i1}
\fmfv{decor.size=0, label=${\scriptstyle 2}$, l.dist=1mm, l.angle=-180}{i2}
\end{fmfgraph*}
\end{center}} \hspace*{0.4cm} ,  \hspace*{1cm}
\setlength{\unitlength}{1mm}
\ddphi{
\setlength{\unitlength}{1mm}
\parbox{10mm}{\centerline{
  \begin{fmfgraph*}(8,6)
  \setval
  \fmfforce{0w,1/2h}{v1}
  \fmfforce{1w,1/2h}{v2}
  \fmfforce{1/2w,1h}{v3}
  \fmfforce{1/2w,0h}{v4}
  \fmfforce{1/2w,1/2h}{v5}
  \fmf{plain,left=0.4}{v1,v3,v2,v4,v1}
  \fmfv{decor.size=0, label=${\scriptstyle 1}$, l.dist=0mm, l.angle=0}{v5}
  \end{fmfgraph*} } } 
} =  \hspace*{0.2cm}
\setlength{\unitlength}{1mm}
\parbox{5mm}{\begin{center}
\begin{fmfgraph*}(3,9)
\setval
\fmfstraight
\fmfforce{1w,0.835h}{o1}
\fmfforce{1w,0.165h}{o2}
\fmfforce{0w,1h}{i1}
\fmfforce{0w,0.66h}{i2}
\fmfforce{0w,0.33h}{i3}
\fmfforce{0w,0h}{i4}
\fmf{dashes}{o1,o2}
\fmf{fermion}{i1,o1}
\fmf{fermion}{o1,i2}
\fmf{fermion}{i3,o2}
\fmf{fermion}{o2,i4}
\fmfdot{o1,o2}
\fmfv{decor.size=0, label=${\scriptstyle 1}$, l.dist=1mm, l.angle=-180}{i1}
\fmfv{decor.size=0, label=${\scriptstyle 2}$, l.dist=1mm, l.angle=-180}{i2}
\fmfv{decor.size=0, label=${\scriptstyle 3}$, l.dist=1mm, l.angle=-180}{i3}
\fmfv{decor.size=0, label=${\scriptstyle 4}$, l.dist=1mm, l.angle=-180}{i4}
\end{fmfgraph*}
\end{center}}
\pm\quad 
\setlength{\unitlength}{1mm}
\parbox{5mm}{\begin{center}
\begin{fmfgraph*}(3,9)
\setval
\fmfstraight
\fmfforce{1w,0.835h}{o1}
\fmfforce{1w,0.165h}{o2}
\fmfforce{0w,1h}{i1}
\fmfforce{0w,0.66h}{i2}
\fmfforce{0w,0.33h}{i3}
\fmfforce{0w,0h}{i4}
\fmf{dashes}{o1,o2}
\fmf{fermion}{i1,o1}
\fmf{fermion}{o1,i2}
\fmf{fermion}{i3,o2}
\fmf{fermion}{o2,i4}
\fmfdot{o1,o2}
\fmfv{decor.size=0, label=${\scriptstyle 1}$, l.dist=1mm, l.angle=-180}{i1}
\fmfv{decor.size=0, label=${\scriptstyle 4}$, l.dist=1mm, l.angle=-180}{i2}
\fmfv{decor.size=0, label=${\scriptstyle 3}$, l.dist=1mm, l.angle=-180}{i3}
\fmfv{decor.size=0, label=${\scriptstyle 2}$, l.dist=1mm, l.angle=-180}{i4}
\end{fmfgraph*}
\end{center}}
\hspace*{2mm}.
\label{O3}
\end{eqnarray}
Inserting (\ref{O2}) into (\ref{FUNC7}), where we 
have to take care of connecting 
only legs with the same label, we find the second-order correction of the
grand-canonical potential $\Omega$:
\begin{eqnarray}
\label{O4}
\setlength{\unitlength}{1mm}
\parbox{10mm}{\centerline{
  \begin{fmfgraph*}(8,6)
  \setval
  \fmfforce{0w,1/2h}{v1}
  \fmfforce{1w,1/2h}{v2}
  \fmfforce{1/2w,1h}{v3}
  \fmfforce{1/2w,0h}{v4}
  \fmfforce{1/2w,1/2h}{v5}
  \fmf{plain,left=0.4}{v1,v3,v2,v4,v1}
  \fmfv{decor.size=0, label=${\scriptstyle 2}$, l.dist=0mm, l.angle=0}{v5}
  \end{fmfgraph*} } } 
 = 
%
%
\frac{1}{4}
\setlength{\unitlength}{1mm}
\parbox{18mm}{\begin{center}
\begin{fmfgraph}(8,5)
\setval
\fmfleft{i2,i1}
\fmfright{o2,o1}
\fmf{fermion,left=1}{i1,i2,i1}
\fmf{dashes}{i1,o1}
\fmf{dashes}{i2,o2}
\fmf{fermion,left=1}{o1,o2,o1}
\fmfdot{i1,i2,o1,o2}
\end{fmfgraph}
\end{center}}
+
%
%
\setlength{\unitlength}{1mm}
\parbox{18mm}{\begin{center}
\begin{fmfgraph}(13,5)
\setval
\fmfforce{0.385w,0.5h}{v1}
\fmfforce{0.615w,0.5h}{v2}
\fmfforce{0.192w,1h}{v3}
\fmfforce{0.192w,0h}{v4}
\fmf{fermion,left=0.5}{v3,v1,v4}
\fmf{fermion,left=1}{v4,v3}
\fmf{dashes}{v3,v4}
\fmf{dashes}{v1,v2}
\fmfi{fermion}{fullcircle rotated 180 scaled 0.384w shifted (0.808w,0.5h)}
\fmfdot{v1,v2,v3,v4}
\end{fmfgraph}
\end{center}}
%
%
\pm\,\frac{1}{2}
\setlength{\unitlength}{1mm}
\parbox{26mm}{\begin{center}
\begin{fmfgraph}(21,5)
\setval
\fmfforce{0.238w,0.5h}{v1}
\fmfforce{0.381w,0.5h}{v2}
\fmfforce{0.619w,0.5h}{v3}
\fmfforce{0.762w,0.5h}{v4}
\fmf{fermion,left=1}{v2,v3,v2}
\fmf{dashes}{v1,v2}
\fmf{dashes}{v3,v4}
\fmfi{fermion}{reverse fullcircle scaled 0.238w shifted (0.119w,0.5h)}
\fmfi{fermion}{fullcircle rotated 180 scaled 0.238w shifted (0.881w,0.5h)}
\fmfdot{v1,v2,v3,v4}
\end{fmfgraph}
\end{center}}
%
%
\pm\,\frac{1}{4}
\setlength{\unitlength}{1mm}
\parbox{10mm}{\begin{center}
\begin{fmfgraph}(5,5)
\setval
\fmfleft{i2,i1}
\fmfright{o2,o1}
\fmf{fermion,left=0.5}{i1,o1,o2,i2,i1}
\fmf{dashes}{i1,o2}
\fmf{dashes}{i2,o1}
\fmfdot{i1,i2,o1,o2}
\end{fmfgraph}
\end{center}}
%
%
\pm\,\frac{1}{2}
\setlength{\unitlength}{1mm}
\parbox{10mm}{\begin{center}
\begin{fmfgraph}(5,5)
\setval
\fmfleft{i2,i1}
\fmfright{o2,o1}
\fmf{fermion,left=0.5}{i1,o1,o2,i2,i1}
\fmf{dashes,left=0.4}{i1,i2}
\fmf{dashes,right=0.4}{o1,o2}
\fmfdot{i1,i2,o1,o2}
\end{fmfgraph}
\end{center}}. 
\end{eqnarray}
The calculation of the third-order correction $\Omega^{(3)}$ 
leads to the following 20 diagrams:
\begin{eqnarray}
\label{O5}
\setlength{\unitlength}{1mm}
\parbox{10mm}{\centerline{
  \begin{fmfgraph*}(8,6)
  \setval
  \fmfforce{0w,1/2h}{v1}
  \fmfforce{1w,1/2h}{v2}
  \fmfforce{1/2w,1h}{v3}
  \fmfforce{1/2w,0h}{v4}
  \fmfforce{1/2w,1/2h}{v5}
  \fmf{plain,left=0.4}{v1,v3,v2,v4,v1}
  \fmfv{decor.size=0, label=${\scriptstyle 3}$, l.dist=0mm, l.angle=0}{v5}
  \end{fmfgraph*} } } 
 &=&  \,\frac{1}{2}
\setlength{\unitlength}{1mm}
\parbox{18mm}{\begin{center}
\begin{fmfgraph}(13,5)
\setval
\fmfforce{0w,0.5h}{i1}
\fmfforce{0.192w,1h}{i2}
\fmfforce{0.385w,0.5h}{i3}
\fmfforce{0.192w,0h}{i4}
\fmfforce{0.808w,1h}{o1}
\fmfforce{0.808w,0h}{o2}
\fmf{fermion,left=0.5}{i1,i2,i3,i4,i1}
\fmf{dashes}{i1,i3}
\fmf{dashes}{i2,o1}
\fmf{dashes}{i4,o2}
\fmf{fermion,left=1}{o1,o2,o1}
\fmfdot{i1,i2,i3,i4,o1,o2}
\end{fmfgraph}
\end{center}}
+\,\frac{1}{6}
\setlength{\unitlength}{1mm}
\parbox{18mm}{\begin{center}
\begin{fmfgraph}(13,5)
\setval
\fmfforce{0.192w,1h}{i1}
\fmfforce{0.385w,0.5h}{i2}
\fmfforce{0.192w,0h}{i3}
\fmfforce{0.808w,1h}{o1}
\fmfforce{0.808w,0h}{o2}
\fmfforce{0.615w,0.5h}{o3}
\fmf{fermion,left=0.5}{i1,i2,i3}
\fmf{fermion,left=1}{i3,i1}
\fmf{dashes}{i1,o1}
\fmf{dashes}{i2,o3}
\fmf{dashes}{i3,o2}
\fmf{fermion,left=1}{o1,o2}
\fmf{fermion,left=0.5}{o2,o3,o1}
\fmfdot{i1,i2,i3,o1,o2,o3}
\end{fmfgraph}
\end{center}}
+\,\frac{1}{6}
\setlength{\unitlength}{1mm}
\parbox{18mm}{\begin{center}
\begin{fmfgraph}(13,5)
\setval
\fmfforce{0.192w,1h}{i1}
\fmfforce{0.385w,0.5h}{i2}
\fmfforce{0.192w,0h}{i3}
\fmfforce{0.808w,1h}{o1}
\fmfforce{0.808w,0h}{o2}
\fmfforce{0.615w,0.5h}{o3}
\fmf{fermion,left=0.5}{i1,i2,i3}
\fmf{fermion,left=1}{i3,i1}
\fmf{dashes}{i1,o1}
\fmf{dashes}{i2,o3}
\fmf{dashes}{i3,o2}
\fmf{fermion,right=1}{o2,o1}
\fmf{fermion,right=0.5}{o1,o3,o2}
\fmfdot{i1,i2,i3,o1,o2,o3}
\end{fmfgraph}
\end{center}}
+
\setlength{\unitlength}{1mm}
\parbox{18mm}{\begin{center}
\begin{fmfgraph}(13,5)
\setval
\fmfforce{0.192w,1h}{i1}
\fmfforce{0.192w,0h}{i2}
\fmfforce{0.026w,0.75h}{i3}
\fmfforce{0.026w,0.25h}{i4}
\fmfforce{0.808w,1h}{o1}
\fmfforce{0.808w,0h}{o2}
\fmf{fermion,left=1}{i1,i2}
\fmf{fermion,left=1/3}{i2,i4,i3,i1}
\fmf{dashes,left=0.8}{i3,i4}
\fmf{dashes}{i1,o1}
\fmf{dashes}{i2,o2}
\fmf{fermion,left=1}{o1,o2,o1}
\fmfdot{i1,i2,i3,i4,o1,o2}
\end{fmfgraph}
\end{center}}
\pm\,\frac{1}{6}
\setlength{\unitlength}{1mm}
\parbox{20mm}{\begin{center}
\begin{fmfgraph}(10,5)
\setval
\fmfforce{0w,1h}{i1}
\fmfforce{0w,0h}{i2}
\fmfforce{0.33w,0h}{v1}
\fmfforce{0.66w,0h}{v2}
\fmfforce{1w,1h}{o1}
\fmfforce{1w,0h}{o2}
\fmf{fermion,left=1}{i1,i2,i1}
\fmf{fermion,left=1}{o1,o2,o1}
\fmf{dashes}{i1,o1}
\fmf{dashes}{i2,v1}
\fmf{fermion,left=1}{v1,v2,v1}
\fmf{dashes}{v2,o2}
\fmfdot{i1,i2,v1,v2,o1,o2}
\end{fmfgraph}
\end{center}}
\nonumber\\ && 
+\,\frac{1}{2}
\setlength{\unitlength}{1mm}
\parbox{18mm}{\begin{center}
\begin{fmfgraph}(13,5)
\setval
\fmfforce{0.192w,1h}{i1}
\fmfforce{0.385w,0.5h}{i2}
\fmfforce{0.192w,0h}{i3}
\fmfforce{0.808w,1h}{o1}
\fmfforce{0.808w,0h}{o2}
\fmfforce{0.615w,0.5h}{o3}
\fmf{fermion,left=0.5}{i1,i2,i3}
\fmf{fermion,left=1}{i3,i1}
\fmf{dashes}{i1,i3}
\fmf{dashes}{i2,o3}
\fmf{dashes}{o1,o2}
\fmf{fermion,right=1}{o2,o1}
\fmf{fermion,right=0.5}{o1,o3,o2}
\fmfdot{i1,i2,i3,o1,o2,o3}
\end{fmfgraph}
\end{center}}
+
\setlength{\unitlength}{1mm}
\parbox{17mm}{\begin{center}
\begin{fmfgraph}(13,5)
\setval
\fmfforce{0.252w,0.976h}{i2}
\fmfforce{0.037w,0.794h}{i3}
\fmfforce{0.037w,0.206h}{i4}
\fmfforce{0.252w,0.024h}{i5}
\fmfforce{0.385w,0.5h}{i1}
\fmfforce{0.615w,0.5h}{o1}
\fmf{fermion,right=0.3}{i1,i2,i3,i4,i5,i1}
\fmf{dashes}{i3,i5}
\fmf{dashes}{i2,i4}
\fmf{dashes}{i1,o1}
\fmfi{fermion}{fullcircle rotated 180 scaled 0.385w shifted (0.808w,0.5h)}
\fmfdot{i1,i2,i3,i4,i5,o1}
\end{fmfgraph}
\end{center}}
+
\setlength{\unitlength}{1mm}
\parbox{17mm}{\begin{center}
\begin{fmfgraph}(13,5)
\setval
\fmfforce{0.252w,0.976h}{i2}
\fmfforce{0.037w,0.794h}{i3}
\fmfforce{0.037w,0.206h}{i4}
\fmfforce{0.252w,0.024h}{i5}
\fmfforce{0.385w,0.5h}{i1}
\fmfforce{0.615w,0.5h}{o1}
\fmf{fermion,right=0.3}{i1,i2,i3,i4,i5,i1}
\fmf{dashes}{i2,i5}
\fmf{dashes,left=0.7}{i3,i4}
\fmf{dashes}{i1,o1}
\fmfi{fermion}{fullcircle rotated 180 scaled 0.385w shifted (0.808w,0.5h)}
\fmfdot{i1,i2,i3,i4,i5,o1}
\end{fmfgraph}
\end{center}}
+
\setlength{\unitlength}{1mm}
\parbox{17mm}{\begin{center}
\begin{fmfgraph}(13,5)
\setval
\fmfforce{0.252w,0.976h}{i2}
\fmfforce{0.037w,0.794h}{i3}
\fmfforce{0.037w,0.206h}{i4}
\fmfforce{0.252w,0.024h}{i5}
\fmfforce{0.385w,0.5h}{i1}
\fmfforce{0.615w,0.5h}{o1}
\fmf{fermion,right=0.3}{i1,i2,i3,i4,i5,i1}
\fmf{dashes,left=0.7}{i2,i3}
\fmf{dashes,left=0.7}{i4,i5}
\fmf{dashes}{i1,o1}
\fmfi{fermion}{fullcircle rotated 180 scaled 0.385w shifted (0.808w,0.5h)}
\fmfdot{i1,i2,i3,i4,i5,o1}
\end{fmfgraph}
\end{center}}
\pm
\setlength{\unitlength}{1mm}
\parbox{25mm}{\begin{center}
\begin{fmfgraph}(21,5)
\setval
\fmfforce{0.238w,0.5h}{i1}
\fmfforce{0.381w,0.5h}{v1}
\fmfforce{0.5w,1h}{v2}
\fmfforce{0.5w,0h}{v3}
\fmfforce{0.881w,1h}{o1}
\fmfforce{0.881w,0h}{o2}
\fmfi{fermion}{reverse fullcircle scaled 0.238w shifted (0.119w,0.5h)}
\fmf{dashes}{i1,v1}
\fmf{fermion,right=0.5}{v2,v1,v3}
\fmf{fermion,right=1}{v3,v2}
\fmf{dashes}{v2,o1}
\fmf{dashes}{v3,o2}
\fmf{fermion,right=1}{o1,o2,o1}
\fmfdot{i1,v1,v2,v3,o1,o2}
\end{fmfgraph}
\end{center}}
\nonumber\\ &&  
\pm\,\frac{1}{2}
\setlength{\unitlength}{1mm}
\parbox{25mm}{\begin{center}
\begin{fmfgraph}(21,5)
\setval
\fmfforce{0.238w,0.5h}{v1}
\fmfforce{0.381w,0.5h}{v2}
\fmfforce{0.5w,1h}{v3}
\fmfforce{0.5w,0h}{v4}
\fmfforce{0.619w,0.5h}{v5}
\fmfforce{0.762w,0.5h}{v6}
\fmfi{fermion}{reverse fullcircle scaled 0.238w shifted (0.119w,0.5h)}
\fmfi{fermion}{fullcircle rotated 180 scaled 0.238w shifted (0.881w,0.5h)}
\fmf{dashes}{v1,v2}
\fmf{fermion,left=0.5}{v2,v3,v5,v4,v2}
\fmf{dashes}{v3,v4}
\fmf{dashes}{v5,v6}
\fmfdot{v1,v2,v3,v4,v5,v6}
\end{fmfgraph}
\end{center}}
\pm
\setlength{\unitlength}{1mm}
\parbox{25mm}{\begin{center}
\begin{fmfgraph}(21,5)
\setval
\fmfforce{0.238w,0.5h}{v1}
\fmfforce{0.381w,0.5h}{v2}
\fmfforce{0.5595w,0.933h}{v4}
\fmfforce{0.4405w,0.933h}{v3}
\fmfforce{0.619w,0.5h}{v5}
\fmfforce{0.762w,0.5h}{v6}
\fmfi{fermion}{reverse fullcircle scaled 0.238w shifted (0.119w,0.5h)}
\fmfi{fermion}{fullcircle rotated 180 scaled 0.238w shifted (0.881w,0.5h)}
\fmf{dashes}{v1,v2}
\fmf{fermion,left=0.33}{v2,v3,v4,v5}
\fmf{fermion,left=1}{v5,v2}
\fmf{dashes,right=0.8}{v3,v4}
\fmf{dashes}{v5,v6}
\fmfdot{v1,v2,v3,v4,v5,v6}
\end{fmfgraph}
\end{center}}
\pm
\setlength{\unitlength}{1mm}
\parbox{25mm}{\begin{center}
\begin{fmfgraph}(21,5)
\setval
\fmfforce{0.238w,0.5h}{v1}
\fmfforce{0.381w,0.5h}{v2}
\fmfforce{0.119w,1h}{v4}
\fmfforce{0.119w,0h}{v3}
\fmfforce{0.619w,0.5h}{v5}
\fmfforce{0.762w,0.5h}{v6}
\fmfi{fermion}{reverse fullcircle scaled 0.238w shifted (0.119w,0.5h)}
\fmfi{fermion}{fullcircle rotated 180 scaled 0.238w shifted (0.881w,0.5h)}
\fmf{dashes}{v1,v2}
\fmf{fermion,left=1}{v2,v5,v2}
\fmf{dashes}{v3,v4}
\fmf{dashes}{v5,v6}
\fmfdot{v1,v2,v3,v4,v5,v6}
\end{fmfgraph}
\end{center}}
+\,\frac{1}{3}
\setlength{\unitlength}{1mm}
\parbox{25mm}{\begin{center}
\begin{fmfgraph}(21,13)
\setval
\fmfforce{0.238w,0.192h}{v1}
\fmfforce{0.381w,0.192h}{v2}
\fmfforce{0.5w,0.384h}{v4}
\fmfforce{0.5w,0.615h}{v3}
\fmfforce{0.619w,0.192h}{v5}
\fmfforce{0.762w,0.192h}{v6}
\fmfi{fermion}{reverse fullcircle scaled 0.238w shifted (0.119w,0.192h)}
\fmfi{fermion}{fullcircle rotated 180 scaled 0.238w shifted (0.881w,0.192h)}
\fmfi{fermion}{fullcircle rotated 270 scaled 0.238w shifted (0.5w,0.808h)}
\fmf{dashes}{v1,v2}
\fmf{fermion,left=0.5}{v2,v4,v5}
\fmf{fermion,left=1}{v5,v2}
\fmf{dashes}{v3,v4}
\fmf{dashes}{v5,v6}
\fmfdot{v1,v2,v3,v4,v5,v6}
\end{fmfgraph}
\end{center}}
\nonumber\\ && 
+\,\frac{1}{2}
\setlength{\unitlength}{1mm}
\parbox{32mm}{\begin{center}
\begin{fmfgraph}(29,5)
\setval
\fmfforce{0.172w,0.5h}{v1}
\fmfforce{0.276w,0.5h}{v2}
\fmfforce{0.448w,0.5h}{v3}
\fmfforce{0.552w,0.5h}{v4}
\fmfforce{0.724w,0.5h}{v5}
\fmfforce{0.828w,0.5h}{v6}
\fmfi{fermion}{reverse fullcircle scaled 0.172w shifted (0.086w,0.5h)}
\fmfi{fermion}{fullcircle rotated 180 scaled 0.172w shifted (0.914w,0.5h)}
\fmf{dashes}{v1,v2}
\fmf{fermion,left=1}{v2,v3,v2}
\fmf{fermion,left=1}{v4,v5,v4}
\fmf{dashes}{v3,v4}
\fmf{dashes}{v5,v6}
\fmfdot{v1,v2,v3,v4,v5,v6}
\end{fmfgraph}
\end{center}}
\pm\,\frac{1}{2}
\setlength{\unitlength}{1mm}
\parbox{10mm}{\begin{center}
\begin{fmfgraph}(7,7)
\setval
\fmfforce{0w,0.5h}{v1}
\fmfforce{0.25w,0.933h}{v2}
\fmfforce{0.75w,0.933h}{v3}
\fmfforce{1w,0.5h}{v4}
\fmfforce{0.75w,0.067h}{v5}
\fmfforce{0.25w,0.067h}{v6}
\fmf{fermion,left=0.3}{v1,v2,v3,v4,v5,v6,v1}
\fmf{dashes}{v1,v4}
\fmf{dashes}{v2,v6}
\fmf{dashes}{v3,v5}
\fmfdot{v1,v2,v3,v4,v5,v6}
\end{fmfgraph}
\end{center}}
\pm\,\frac{1}{6}
\setlength{\unitlength}{1mm}
\parbox{10mm}{\begin{center}
\begin{fmfgraph}(7,7)
\setval
\fmfforce{0w,0.5h}{v1}
\fmfforce{0.25w,0.933h}{v2}
\fmfforce{0.75w,0.933h}{v3}
\fmfforce{1w,0.5h}{v4}
\fmfforce{0.75w,0.067h}{v5}
\fmfforce{0.25w,0.067h}{v6}
\fmf{fermion,left=0.3}{v1,v2,v3,v4,v5,v6,v1}
\fmf{dashes}{v1,v4}
\fmf{dashes}{v2,v5}
\fmf{dashes}{v3,v6}
\fmfdot{v1,v2,v3,v4,v5,v6}
\end{fmfgraph}
\end{center}}
\pm
\setlength{\unitlength}{1mm}
\parbox{10mm}{\begin{center}
\begin{fmfgraph}(7,7)
\setval
\fmfforce{0w,0.5h}{v1}
\fmfforce{0.25w,0.933h}{v2}
\fmfforce{0.75w,0.933h}{v3}
\fmfforce{1w,0.5h}{v4}
\fmfforce{0.75w,0.067h}{v5}
\fmfforce{0.25w,0.067h}{v6}
\fmf{fermion,left=0.3}{v1,v2,v3,v4,v5,v6,v1}
\fmf{dashes,right=0.7}{v2,v3}
\fmf{dashes,right=0.2}{v4,v6}
\fmf{dashes,right=0.2}{v5,v1}
\fmfdot{v1,v2,v3,v4,v5,v6}
\end{fmfgraph}
\end{center}}
\pm\,\frac{1}{2}
\setlength{\unitlength}{1mm}
\parbox{10mm}{\begin{center}
\begin{fmfgraph}(7,7)
\setval
\fmfforce{0w,0.5h}{v1}
\fmfforce{0.25w,0.933h}{v2}
\fmfforce{0.75w,0.933h}{v3}
\fmfforce{1w,0.5h}{v4}
\fmfforce{0.75w,0.067h}{v5}
\fmfforce{0.25w,0.067h}{v6}
\fmf{fermion,left=0.3}{v1,v2,v3,v4,v5,v6,v1}
\fmf{dashes,right=0.7}{v2,v3}
\fmf{dashes}{v1,v4}
\fmf{dashes,right=0.7}{v5,v6}
\fmfdot{v1,v2,v3,v4,v5,v6}
\end{fmfgraph}
\end{center}}
\pm\,\frac{1}{3}
\setlength{\unitlength}{1mm}
\parbox{10mm}{\begin{center}
\begin{fmfgraph}(7,7)
\setval
\fmfforce{0w,0.5h}{v1}
\fmfforce{0.25w,0.933h}{v2}
\fmfforce{0.75w,0.933h}{v3}
\fmfforce{1w,0.5h}{v4}
\fmfforce{0.75w,0.067h}{v5}
\fmfforce{0.25w,0.067h}{v6}
\fmf{fermion,left=0.3}{v1,v2,v3,v4,v5,v6,v1}
\fmf{dashes,right=0.7}{v2,v3}
\fmf{dashes,right=0.7}{v4,v5}
\fmf{dashes,right=0.7}{v6,v1}
\fmfdot{v1,v2,v3,v4,v5,v6}
\end{fmfgraph}
\end{center}}
\hspace*{2mm} . 
\end{eqnarray}
From the vacuum diagrams (\ref{BEG}), (\ref{O4}), and (\ref{O5}),
we observe a simple mnemonic rule 
for the weights of the connected vacuum diagrams which is similar to a corresponding one for QED \cite{QED,SDQED1}. 
At least up to four loops, each weight is
equal to the reciprocal number of lines, whose amputation leads to the same two-point diagram.
The sign is given by $(\pm)^l$, where $l$ denotes the number of 
loops. Let us also note that the sum of all weights of the connected 
vacuum diagrams in the loop order under consideration vanishes. 
These simple weights are a consequence of the bose and fermi statistics as well as  
the interaction (\ref{VE}). The weights of the vacuum diagrams in other  
theories, like $\phi^4$-theory~\cite{SYM,ASYM1,ASYM2,Neu,Verena}, 
follow more complicated rules.
\section{Hugenholtz Diagrams}\label{HuGo}
Now we consider the special case of a dilute bose gas where the two-particle interaction $V^{({\rm int})}$ can be 
approximated by the local one
\begin{eqnarray}
V^{({\rm int})}_{1234} = \frac{2 \lambda^2}{\beta} \, a \, \delta_{12} \delta_{13} \delta_{14} \, .
\end{eqnarray}
Here $\lambda$ denotes the thermal wave length and $a$ the scattering length. This specification modifies the
graphical representation (\ref{VE}) of the two-particle interaction. The dashed line is removed and the
two vertices are united to a single one:
\begin{eqnarray}
\setlength{\unitlength}{1mm}
\parbox{12mm}{\centerline{
\begin{fmfgraph*}(10,5)
\setval
\fmfforce{0w,0h}{i1}
\fmfforce{0w,1h}{i2}
\fmfforce{1w,0h}{o1}
\fmfforce{1w,1h}{o2}
\fmfforce{3/10w,0.5h}{v1}
\fmfforce{7/10w,0.5h}{v2}
\fmfforce{3/10w,1/10h}{v3}
\fmfforce{7/10w,1/10h}{v4}
\fmf{dashes}{v1,v2}
\fmf{electron}{v1,i1}
\fmf{electron}{i2,v1}
\fmf{electron}{o1,v2}
\fmf{electron}{v2,o2}
\fmfdot{v1,v2}
\fmfv{decor.size=0, label=${\scriptstyle 1}$, l.dist=1mm, l.angle=180}{i1}
\fmfv{decor.size=0, label=${\scriptstyle 2}$, l.dist=1mm, l.angle=180}{i2}
\fmfv{decor.size=0, label=${\scriptstyle 3}$, l.dist=1mm, l.angle=0}{o2}
\fmfv{decor.size=0, label=${\scriptstyle 4}$, l.dist=1mm, l.angle=0}{o1}
\end{fmfgraph*}}} 
\hspace*{0.5cm} \Longrightarrow \hspace*{0.5cm}
\setlength{\unitlength}{1mm}
\parbox{11mm}{\begin{center}
\begin{fmfgraph*}(5,5)
\setval
\fmfstraight
\fmfforce{0w,0h}{o2}
\fmfforce{0w,1h}{i1}
\fmfforce{1w,0h}{o1}
\fmfforce{1w,1h}{i2}
\fmfforce{1/2w,1/2h}{v1}
\fmf{fermion}{i1,v1}
\fmf{fermion}{v1,i2}
\fmf{fermion}{o1,v1}
\fmf{fermion}{v1,o2}
\fmfdot{v1}
\fmfv{decor.size=0, label=${\scriptstyle 2}$, l.dist=1mm, l.angle=180}{i1}
\fmfv{decor.size=0, label=${\scriptstyle 3}$, l.dist=1mm, l.angle=0}{i2}
\fmfv{decor.size=0, label=${\scriptstyle 1}$, l.dist=1mm, l.angle=180}{o2}
\fmfv{decor.size=0, label=${\scriptstyle 4}$, l.dist=1mm, l.angle=0}{o1}
\end{fmfgraph*}
\end{center}} 
\hspace*{0.5cm}.
\label{RED}
\end{eqnarray}
The resulting vacuum diagrams are called Hugenholtz diagrams \cite{Negele,Hugenholtz}. 
In this way the direct and the exchange vacuum
diagram of the first order (\ref{BEG}) of the grand-canonical
potential reduce to 
\begin{eqnarray}
\label{NBEG}
\setlength{\unitlength}{1mm}
\parbox{10mm}{\centerline{
  \begin{fmfgraph*}(8,6)
  \setval
  \fmfforce{0w,1/2h}{v1}
  \fmfforce{1w,1/2h}{v2}
  \fmfforce{1/2w,1h}{v3}
  \fmfforce{1/2w,0h}{v4}
  \fmfforce{1/2w,1/2h}{v5}
  \fmf{plain,left=0.4}{v1,v3,v2,v4,v1}
  \fmfv{decor.size=0, label=${\scriptstyle 1}$, l.dist=0mm, l.angle=0}{v5}
  \end{fmfgraph*} } }
\hspace*{0.2cm} = \hspace*{0.2cm}
\setlength{\unitlength}{1mm}
\parbox{8mm}{\begin{center}
\begin{fmfgraph}(10,5)
\setval
\fmfforce{1/4w,1h}{o1}
\fmfforce{1/4w,0h}{u1}
\fmfforce{1/2w,1/2h}{v2}
\fmfforce{3/4w,1h}{o3}
\fmfforce{3/4w,0h}{u3}
\fmf{fermion,right=1}{o1,u1}
\fmf{fermion,right=1}{u3,o3}
\fmf{plain,right=1}{o3,u3}
\fmf{plain,right=1}{u1,o1}
\fmfdot{v2}
\end{fmfgraph}\end{center}}
\end{eqnarray}
which is also shown as diagram \# 1.1 in Table \ref{resu}.
Correspondingly, the second and the third order (\ref{O4}) and (\ref{O5})  lead to the vacuum diagrams depicted in Table \ref{resu}.
We observe that the weight of a Hugenholtz diagram for the weakly interacting bose gas is given by the formula
\begin{eqnarray}
\label{WEI}
W = \frac{2^p}{2^{D+T} 4^F N}
\end{eqnarray}
(compare with the corresponding formula for the $\phi^4$-theory \cite{SYM,Neu,Verena}).
Here $D,T,F$ denote the number of the following double, triple and fourfold connections between two vertices:
\begin{eqnarray}
D: \hspace*{0.2cm}
\setlength{\unitlength}{1mm}
\parbox{14mm}{\begin{center}
\begin{fmfgraph}(10,5)
\setval
\fmfstraight
\fmfforce{0w,0h}{i1}
\fmfforce{0w,1h}{i2}
\fmfforce{1w,0h}{o1}
\fmfforce{1w,1h}{o2}
\fmfforce{2.5/10w,0.5h}{v1}
\fmfforce{7.5/10w,0.5h}{v2}
\fmf{fermion}{v1,i1}
\fmf{fermion}{o1,v2}
\fmf{fermion}{v1,i2}
\fmf{fermion}{o2,v2}
\fmf{fermion,right=1}{v2,v1}
\fmf{fermion,left=1}{v2,v1}
\fmfdot{v1,v2}
\end{fmfgraph}
\end{center}}
\, , \hspace*{1cm} T: 
\setlength{\unitlength}{1mm}
\parbox{18mm}{\begin{center}
\begin{fmfgraph}(11,5)
\setval
\fmfforce{0w,0.5h}{v1}
\fmfforce{3/11w,0.5h}{v2}
\fmfforce{8/11w,0.5h}{v3}
\fmfforce{1w,0.5h}{v4}
\fmf{fermion}{v2,v1}
\fmf{fermion}{v3,v2}
\fmf{fermion}{v4,v3}
\fmf{fermion,right=1}{v2,v3}
\fmf{fermion,right=1}{v3,v2}
\fmfdot{v2,v3}
\end{fmfgraph}  \end{center}}
\, , \hspace*{1cm} F: \hspace*{0.2cm}
\parbox{7.5mm}{\begin{center}
\begin{fmfgraph}(7.5,7.5)
\setval
\fmfforce{0w,0.5h}{v1}
\fmfforce{1w,0.5h}{v2}
\fmf{fermion,right=1}{v1,v2,v1}
\fmf{fermion,left=0.4}{v1,v2,v1}
\fmfdot{v1,v2}
\end{fmfgraph}\end{center}} 
\hspace*{0.5cm} .
\end{eqnarray}
Furthermore, the number $N$ stands for the number of vertex permutations leaving the vacuum diagram unchanged, where
the vertices remain attached to the lines emerging from them in the same way as before.
The Hugenholtz diagrams also follow from a graphical recursion relation. From (\ref{FUNC7}) and (\ref{RED}) we get 
for $p \ge 1$:
\begin{eqnarray}
  \setlength{\unitlength}{1mm}
\parbox{10mm}{\centerline{
  \begin{fmfgraph*}(9,6.7)
  \setval
  \fmfforce{0w,1/2h}{v1}
  \fmfforce{1w,1/2h}{v2}
  \fmfforce{1/2w,1h}{v3}
  \fmfforce{1/2w,0h}{v4}
  \fmfforce{1/2w,1/2h}{v5}
  \fmf{plain,left=0.4}{v1,v3,v2,v4,v1}
  \fmfv{decor.size=0, label=${\scriptstyle p+1}$, l.dist=0mm, l.angle=0}{v5}
  \end{fmfgraph*} } }  
& = &  \frac{1}{2(p+1)} \hspace*{1mm}\left( 4 \hspace*{0.5mm}
  \setlength{\unitlength}{1mm}
\parbox{12mm}{\centerline{
  \begin{fmfgraph*}(10,6)
  \setval
  \fmfforce{2.5/10w,0.5/6h}{v1b}
  \fmfforce{2.5/10w,5.5/6h}{v2b}
  \fmfforce{5/10w,3/6h}{v2}
  \fmfforce{1w,5.5/6h}{v3}
  \fmfforce{1w,0.5/6h}{v4}
\fmf{plain,right=1}{v1b,v2b}
  \fmf{fermion,right=1}{v2b,v1b}
  \fmf{fermion,left=0.3}{v4,v2}  
  \fmf{fermion,left=0.3}{v2,v3}
  \fmfv{decor.size=0, label=${\scriptstyle 1}$, l.dist=1mm, l.angle=0}{v3}
  \fmfv{decor.size=0, label=${\scriptstyle 2}$, l.dist=1mm, l.angle=0}{v4}
  \fmfdot{v2}
  \end{fmfgraph*} } } 
\hspace*{3mm}
\dphi{
  \setlength{\unitlength}{1mm}
\parbox{10mm}{\centerline{
  \begin{fmfgraph*}(8,6)
  \setval
  \fmfforce{0w,1/2h}{v1}
  \fmfforce{1w,1/2h}{v2}
  \fmfforce{1/2w,1h}{v3}
  \fmfforce{1/2w,0h}{v4}
  \fmfforce{1/2w,1/2h}{v5}
  \fmf{plain,left=0.4}{v1,v3,v2,v4,v1}
  \fmfv{decor.size=0, label=${\scriptstyle p}$, l.dist=0mm, l.angle=0}{v5}
  \end{fmfgraph*} } }  
}{1}{2}
\hspace*{3mm} +  \hspace*{2mm}
  \setlength{\unitlength}{1mm}
\parbox{8mm}{\centerline{
  \begin{fmfgraph*}(5,9)
  \setval
  \fmfforce{0w,5.5/9h}{v1}
  \fmfforce{1w,10/9h}{v2}
  \fmfforce{1w,7/9h}{v3}
  \fmfforce{1w,4/9h}{v4}
  \fmfforce{1w,1/9h}{v5}
  \fmf{fermion,left=0.35}{v1,v2}
  \fmf{fermion,right=0.15}{v3,v1}
  \fmf{fermion,right=0.15}{v1,v4}  
  \fmf{fermion,left=0.35}{v5,v1}
  \fmfv{decor.size=0, label=${\scriptstyle 1}$, l.dist=1mm, l.angle=0}{v2}
  \fmfv{decor.size=0, label=${\scriptstyle 2}$, l.dist=1mm, l.angle=0}{v3}
  \fmfv{decor.size=0, label=${\scriptstyle 3}$, l.dist=1mm, l.angle=0}{v4}
  \fmfv{decor.size=0, label=${\scriptstyle 4}$, l.dist=1mm, l.angle=0}{v5}
  \fmfdot{v1}
  \end{fmfgraph*} } } 
\hspace*{3mm}
\ddphi{
  \setlength{\unitlength}{1mm}
\parbox{10mm}{\centerline{
  \begin{fmfgraph*}(8,6)
  \setval
  \fmfforce{0w,1/2h}{v1}
  \fmfforce{1w,1/2h}{v2}
  \fmfforce{1/2w,1h}{v3}
  \fmfforce{1/2w,0h}{v4}
  \fmfforce{1/2w,1/2h}{v5}
  \fmf{plain,left=0.4}{v1,v3}
 \fmf{plain,left=0.4}{v3,v2}
 \fmf{plain,left=0.4}{v2,v4}
 \fmf{plain,left=0.4}{v4,v1}
  \fmfv{decor.size=0, label=${\scriptstyle p}$, l.dist=0mm, l.angle=0}{v5}
  \end{fmfgraph*} } }  
}
\nonumber  \right.  \\*[6mm]
&  & \left. + \hspace*{2mm} \sum_{q=1}^{p-1} \hspace*{3mm}
\dphi{
  \setlength{\unitlength}{1mm}
\parbox{11mm}{\centerline{
  \begin{fmfgraph*}(9,6.7)
  \setval
  \fmfforce{0w,1/2h}{v1}
  \fmfforce{1w,1/2h}{v2}
  \fmfforce{1/2w,1h}{v3}
  \fmfforce{1/2w,0h}{v4}
  \fmfforce{1/2w,1/2h}{v5}
  \fmf{plain,left=0.4}{v1,v3}
\fmf{plain,left=0.4}{v3,v2}
\fmf{plain,left=0.4}{v2,v4}
\fmf{plain,left=0.4}{v4,v1}
  \fmfv{decor.size=0, label=${\scriptstyle q}$, l.dist=0mm, l.angle=0}{v5}
  \end{fmfgraph*} } }  
}{1}{2}
\hspace*{3mm}
  \setlength{\unitlength}{1mm}
\parbox{12mm}{\centerline{
  \begin{fmfgraph*}(9,6)
  \setval
  \fmfforce{0w,6.5/6h}{v1}
  \fmfforce{0w,0.5/6h}{v2}
  \fmfforce{1/2w,3.5/6h}{v3}
  \fmfforce{1w,6.5/6h}{v4}
  \fmfforce{1w,0.5/6h}{v5}
  \fmf{fermion,right=0.3}{v3,v1}
  \fmf{fermion,right=0.3}{v2,v3}
  \fmf{fermion,right=0.3}{v3,v5}  
  \fmf{fermion,right=0.3}{v4,v3}
  \fmfv{decor.size=0, label=${\scriptstyle 1}$, l.dist=1mm, l.angle=180}{v1}
  \fmfv{decor.size=0, label=${\scriptstyle 2}$, l.dist=1mm, l.angle=180}{v2}
  \fmfv{decor.size=0, label=${\scriptstyle 3}$, l.dist=1mm, l.angle=0}{v5}
  \fmfv{decor.size=0, label=${\scriptstyle 4}$, l.dist=1mm, l.angle=0}{v4}
  \fmfdot{v3}
  \end{fmfgraph*} } } 
\hspace*{3mm}
\dphi{
\setlength{\unitlength}{1mm}
  \parbox{11mm}{\centerline{
  \begin{fmfgraph*}(9,6.7)
  \setval
  \fmfforce{0w,1/2h}{v1}
  \fmfforce{1w,1/2h}{v2}
  \fmfforce{1/2w,1h}{v3}
  \fmfforce{1/2w,0h}{v4}
  \fmfforce{1/2w,1/2h}{v5}
  \fmf{plain,left=0.4}{v1,v3}
\fmf{plain,left=0.4}{v3,v2}
\fmf{plain,left=0.4}{v2,v4}
\fmf{plain,left=0.4}{v4,v1}
  \fmfv{decor.size=0, label=${\scriptstyle p-q}$, l.dist=0mm, l.angle=0}{v5}
  \end{fmfgraph*} } }  
}{3}{4}
\right) \hspace*{5mm} .
\label{HUGOREK}
\end{eqnarray}
This is iterated starting from (\ref{NBEG}).
Indeed, amputating one or two lines in the first order (\ref{NBEG}), i.e.
\begin{eqnarray}
\dphi{
\setlength{\unitlength}{1mm}
\parbox{10mm}{\centerline{
  \begin{fmfgraph*}(8,6)
  \setval
  \fmfforce{0w,1/2h}{v1}
  \fmfforce{1w,1/2h}{v2}
  \fmfforce{1/2w,1h}{v3}
  \fmfforce{1/2w,0h}{v4}
  \fmfforce{1/2w,1/2h}{v5}
  \fmf{plain,left=0.4}{v1,v3,v2,v4,v1}
  \fmfv{decor.size=0, label=${\scriptstyle 1}$, l.dist=0mm, l.angle=0}{v5}
  \end{fmfgraph*} } } 
}{1}{2} = \hspace*{2mm} 2 \hspace*{2mm}
\setlength{\unitlength}{1mm}
\parbox{9mm}{\centerline{
  \begin{fmfgraph*}(4.5,4.5)
  \setval
  \fmfforce{0w,0h}{v1}
  \fmfforce{0w,1h}{v2}
  \fmfforce{5/4.5w,5/4.5h}{v3}
  \fmfforce{5/4.5w,-0.5/4.5h}{v4}
  \fmfforce{1/2w,1/2h}{v5}
  \fmf{fermion}{v5,v1}
  \fmf{fermion}{v2,v5}
  \fmf{plain,right=1}{v3,v4}
\fmf{fermion,left=1}{v3,v4}
  \fmfv{decor.size=0, label=${\scriptstyle 2}$, l.dist=1mm, l.angle=-135}{v1}
  \fmfv{decor.size=0, label=${\scriptstyle 1}$, l.dist=1mm, l.angle=135}{v2}
  \fmfdot{v5}
  \end{fmfgraph*} } } 
\hspace*{0.3cm} , \hspace*{1cm}
\ddphi{
\setlength{\unitlength}{1mm}
\parbox{10mm}{\centerline{
  \begin{fmfgraph*}(8,6)
  \setval
  \fmfforce{0w,1/2h}{v1}
  \fmfforce{1w,1/2h}{v2}
  \fmfforce{1/2w,1h}{v3}
  \fmfforce{1/2w,0h}{v4}
  \fmfforce{1/2w,1/2h}{v5}
  \fmf{plain,left=0.4}{v1,v3,v2,v4,v1}
  \fmfv{decor.size=0, label=${\scriptstyle 1}$, l.dist=0mm, l.angle=0}{v5}
  \end{fmfgraph*} } } 
} = \hspace*{2mm} 2 \hspace*{2mm}
\setlength{\unitlength}{1mm}
\parbox{7mm}{\centerline{
  \begin{fmfgraph*}(4.5,4.5)
  \setval
  \fmfforce{0w,0h}{v1}
  \fmfforce{0w,1h}{v2}
  \fmfforce{1w,1h}{v3}
  \fmfforce{1w,0h}{v4}
\fmfforce{1/2w,1/2h}{vm}
  \fmfforce{1/2w,1/2h}{v5}
  \fmf{fermion}{vm,v1}
  \fmf{fermion}{v2,vm}
  \fmf{fermion}{vm,v3}
  \fmf{fermion}{v4,vm}
\fmfv{decor.size=0, label=${\scriptstyle 2}$, l.dist=1mm, l.angle=-135}{v1}
  \fmfv{decor.size=0, label=${\scriptstyle 1}$, l.dist=1mm, l.angle=135}{v2}
  \fmfv{decor.size=0, label=${\scriptstyle 4}$, l.dist=1mm, l.angle=45}{v3}
  \fmfv{decor.size=0, label=${\scriptstyle 3}$, l.dist=1mm, l.angle=-45}{v4}
  \fmfdot{v5}
  \end{fmfgraph*} } } \hspace*{0.3cm} ,  
\end{eqnarray}
we obtain from (\ref{HUGOREK}) for $p=1$ the second-order diagrams in Table \ref{resu}. A subsequent amputation of one line
\begin{eqnarray}
\dphi{
\setlength{\unitlength}{1mm}
\parbox{10mm}{\centerline{
  \begin{fmfgraph*}(8,6)
  \setval
  \fmfforce{0w,1/2h}{v1}
  \fmfforce{1w,1/2h}{v2}
  \fmfforce{1/2w,1h}{v3}
  \fmfforce{1/2w,0h}{v4}
  \fmfforce{1/2w,1/2h}{v5}
  \fmf{plain,left=0.4}{v1,v3,v2,v4,v1}
  \fmfv{decor.size=0, label=${\scriptstyle 2}$, l.dist=0mm, l.angle=0}{v5}
  \end{fmfgraph*} } } 
}{1}{2} 
\hspace*{0.2cm} = \hspace*{0.2cm} 4 \hspace*{0.1cm}
\setlength{\unitlength}{1mm}
\parbox{20mm}{\begin{center}
\begin{fmfgraph*}(17,5)
\setval
\fmfforce{2/17w,0h}{i1}
\fmfforce{5/17w,0h}{v1}
\fmfforce{2.5/17w,1/2h}{v2a}
\fmfforce{7.5/17w,1/2h}{v2b}
\fmfforce{12/17w,0h}{v3}
\fmfforce{9.5/17w,1/2h}{v4a}
\fmfforce{14.5/17w,1/2h}{v4b}
\fmfforce{15/17w,0h}{o1}
\fmf{fermion}{v1,i1}
\fmf{fermion}{v3,v1}
\fmf{fermion}{o1,v3}
\fmf{fermion,right=1}{v2b,v2a}
\fmf{plain,right=1}{v2a,v2b}
\fmf{fermion,right=1}{v4b,v4a}
\fmf{plain,right=1}{v4a,v4b}
\fmfdot{v1,v3}
\fmfv{decor.size=0, label=${\scriptstyle 2}$, l.dist=1mm, l.angle=180}{i1}
\fmfv{decor.size=0, label=${\scriptstyle 1}$, l.dist=1mm, l.angle=0}{o1}
\end{fmfgraph*} \end{center}}
\hspace*{0.1cm} + \, 4 \hspace*{0.1cm} 
\parbox{13mm}{\begin{center}
\begin{fmfgraph*}(6,10)
\setval
\fmfforce{0w,0h}{v1}
\fmfforce{0.5w,0h}{v2}
\fmfforce{1w,0h}{v3}
\fmfforce{0.5w,0.5h}{v4}
\fmfforce{0.5/6w,3/4h}{v5a}
\fmfforce{5.5/6w,3/4h}{v5b}
\fmf{fermion}{v2,v1}
\fmf{fermion}{v3,v2}
\fmf{fermion,left=1}{v2,v4}
\fmf{fermion,left=1}{v4,v2}
\fmf{plain,left=1}{v5b,v5a}
\fmf{fermion,right=1}{v5b,v5a}
\fmfdot{v2,v4}
\fmfv{decor.size=0, label=${\scriptstyle 2}$, l.dist=1mm, l.angle=180}{v1}
\fmfv{decor.size=0, label=${\scriptstyle 1}$, l.dist=1mm, l.angle=0}{v3}
\end{fmfgraph*} \end{center}}
\hspace*{0.1cm} + \, 2 \hspace*{0.1cm} 
\setlength{\unitlength}{1mm}
\parbox{18mm}{\begin{center}
\begin{fmfgraph*}(11,5)
\setval
\fmfforce{0w,0.5h}{v1}
\fmfforce{3/11w,0.5h}{v2}
\fmfforce{8/11w,0.5h}{v3}
\fmfforce{1w,0.5h}{v4}
\fmf{fermion}{v2,v1}
\fmf{fermion}{v3,v2}
\fmf{fermion}{v4,v3}
\fmf{fermion,left=1}{v3,v2}
\fmf{fermion,left=1}{v2,v3}
\fmfdot{v2,v3}
\fmfv{decor.size=0, label=${\scriptstyle 2}$, l.dist=1mm, l.angle=180}{v1}
\fmfv{decor.size=0, label=${\scriptstyle 1}$, l.dist=1mm, l.angle=0}{v4}
\end{fmfgraph*}  \end{center}}
\end{eqnarray}
and two lines
\begin{eqnarray}
\setlength{\unitlength}{1mm}
\ddphi{
\setlength{\unitlength}{1mm}
\parbox{10mm}{\centerline{
  \begin{fmfgraph*}(8,6)
  \setval
  \fmfforce{0w,1/2h}{v1}
  \fmfforce{1w,1/2h}{v2}
  \fmfforce{1/2w,1h}{v3}
  \fmfforce{1/2w,0h}{v4}
  \fmfforce{1/2w,1/2h}{v5}
  \fmf{plain,left=0.4}{v1,v3,v2,v4,v1}
  \fmfv{decor.size=0, label=${\scriptstyle 2}$, l.dist=0mm, l.angle=0}{v5}
  \end{fmfgraph*} } } 
} & = &  
4 \hspace*{0.2cm}
\setlength{\unitlength}{1mm}
\parbox{17mm}{\begin{center}
\begin{fmfgraph*}(11,6)
\setval
\fmfstraight
\fmfforce{0w,1/2h}{i1}
\fmfforce{3/11w,1h}{i2}
\fmfforce{3/11w,0h}{i3}
\fmfforce{1w,1/2h}{o1}
\fmfforce{8/11w,1/2h}{v1}
\fmfforce{5.5/11w,5.5/6h}{v2a}
\fmfforce{10.5/11w,5.5/6h}{v2b}
\fmfforce{3/11w,1/2h}{v3}
\fmf{fermion}{o1,v1}
\fmf{fermion}{v3,i1}
\fmf{fermion}{i2,v3}
\fmf{fermion}{v3,i3}
\fmf{fermion}{v1,v3}
\fmf{plain,left=1}{v2b,v2a}
\fmf{fermion,right=1}{v2b,v2a}
\fmfdot{v1,v3}
\fmfv{decor.size=0, label=${\scriptstyle 1}$, l.dist=1mm, l.angle=0}{o1}
\fmfv{decor.size=0, label=${\scriptstyle 2}$, l.dist=1mm, l.angle=180}{i1}
\fmfv{decor.size=0, label=${\scriptstyle 3}$, l.dist=1mm, l.angle=90}{i2}
\fmfv{decor.size=0, label=${\scriptstyle 4}$, l.dist=1mm, l.angle=-90}{i3}
\end{fmfgraph*}
\end{center}}
\hspace*{0.2cm} + 4 \hspace*{0.2cm}
\setlength{\unitlength}{1mm}
\parbox{17mm}{\begin{center}
\begin{fmfgraph*}(11,6)
\setval
\fmfstraight
\fmfforce{0w,1/2h}{i1}
\fmfforce{8/11w,1h}{i2}
\fmfforce{8/11w,0h}{i3}
\fmfforce{1w,1/2h}{o1}
\fmfforce{8/11w,1/2h}{v1}
\fmfforce{0.5/11w,5.5/6h}{v2a}
\fmfforce{5.5/11w,5.5/6h}{v2b}
\fmfforce{3/11w,1/2h}{v3}
\fmf{fermion}{o1,v1}
\fmf{fermion}{v3,i1}
\fmf{fermion}{i2,v1}
\fmf{fermion}{v1,i3}
\fmf{fermion}{v1,v3}
\fmf{plain,left=1}{v2b,v2a}
\fmf{fermion,right=1}{v2b,v2a}
\fmfdot{v1,v3}
\fmfv{decor.size=0, label=${\scriptstyle 1}$, l.dist=1mm, l.angle=0}{o1}
\fmfv{decor.size=0, label=${\scriptstyle 2}$, l.dist=1mm, l.angle=180}{i1}
\fmfv{decor.size=0, label=${\scriptstyle 3}$, l.dist=1mm, l.angle=90}{i2}
\fmfv{decor.size=0, label=${\scriptstyle 4}$, l.dist=1mm, l.angle=-90}{i3}
\end{fmfgraph*}
\end{center}}
\hspace*{0.2cm} + 4 \hspace*{0.2cm}
\setlength{\unitlength}{1mm}
\parbox{17mm}{\begin{center}
\begin{fmfgraph*}(11,6)
\setval
\fmfstraight
\fmfforce{0w,1/2h}{i1}
\fmfforce{3/11w,1h}{i2}
\fmfforce{3/11w,0h}{i3}
\fmfforce{1w,1/2h}{o1}
\fmfforce{8/11w,1/2h}{v1}
\fmfforce{5.5/11w,5.5/6h}{v2a}
\fmfforce{10.5/11w,5.5/6h}{v2b}
\fmfforce{3/11w,1/2h}{v3}
\fmf{fermion}{o1,v1}
\fmf{fermion}{v3,i1}
\fmf{fermion}{i2,v3}
\fmf{fermion}{v3,i3}
\fmf{fermion}{v1,v3}
\fmf{plain,left=1}{v2b,v2a}
\fmf{fermion,right=1}{v2b,v2a}
\fmfdot{v1,v3}
\fmfv{decor.size=0, label=${\scriptstyle 3}$, l.dist=1mm, l.angle=0}{o1}
\fmfv{decor.size=0, label=${\scriptstyle 2}$, l.dist=1mm, l.angle=180}{i1}
\fmfv{decor.size=0, label=${\scriptstyle 1}$, l.dist=1mm, l.angle=90}{i2}
\fmfv{decor.size=0, label=${\scriptstyle 4}$, l.dist=1mm, l.angle=-90}{i3}
\end{fmfgraph*}
\end{center}}
\hspace*{0.2cm} + 4 \hspace*{0.2cm}
\setlength{\unitlength}{1mm}
\parbox{17mm}{\begin{center}
\begin{fmfgraph*}(11,6)
\setval
\fmfstraight
\fmfforce{0w,1/2h}{i1}
\fmfforce{8/11w,1h}{i2}
\fmfforce{8/11w,0h}{i3}
\fmfforce{1w,1/2h}{o1}
\fmfforce{8/11w,1/2h}{v1}
\fmfforce{0.5/11w,5.5/6h}{v2a}
\fmfforce{5.5/11w,5.5/6h}{v2b}
\fmfforce{3/11w,1/2h}{v3}
\fmf{fermion}{o1,v1}
\fmf{fermion}{v3,i1}
\fmf{fermion}{i2,v1}
\fmf{fermion}{v1,i3}
\fmf{fermion}{v1,v3}
\fmf{plain,left=1}{v2b,v2a}
\fmf{fermion,right=1}{v2b,v2a}
\fmfdot{v1,v3}
\fmfv{decor.size=0, label=${\scriptstyle 3}$, l.dist=1mm, l.angle=0}{o1}
\fmfv{decor.size=0, label=${\scriptstyle 4}$, l.dist=1mm, l.angle=180}{i1}
\fmfv{decor.size=0, label=${\scriptstyle 1}$, l.dist=1mm, l.angle=90}{i2}
\fmfv{decor.size=0, label=${\scriptstyle 2}$, l.dist=1mm, l.angle=-90}{i3}
\end{fmfgraph*}
\end{center}}
\nonumber \\ && 
+ \, 4 \hspace*{0.2cm}
\setlength{\unitlength}{1mm}
\parbox{8mm}{\begin{center}
\begin{fmfgraph*}(6,5)
\setval
\fmfstraight
\fmfforce{0w,0h}{v1}
\fmfforce{0.5w,0h}{v3}
\fmfforce{1w,0h}{v2}
\fmfforce{0.5/6w,1/2h}{v4a}
\fmfforce{5.5/6w,1/2h}{v4b}
\fmf{fermion}{v3,v1}
\fmf{fermion}{v2,v3}
\fmf{plain,left=1}{v4b,v4a}
\fmf{fermion,right=1}{v4b,v4a}
\fmfdot{v3}
\fmfv{decor.size=0, label=${\scriptstyle 3}$, l.dist=1mm, l.angle=0}{v2}
\fmfv{decor.size=0, label=${\scriptstyle 2}$, l.dist=1mm, l.angle=180}{v1}
\end{fmfgraph*}  \end{center}}
\hspace*{0.4cm}
\setlength{\unitlength}{1mm}
\parbox{8mm}{\begin{center}
\begin{fmfgraph*}(6,5)
\setval
\fmfstraight
\fmfforce{0w,0h}{v1}
\fmfforce{0.5w,0h}{v3}
\fmfforce{1w,0h}{v2}
\fmfforce{0.5/6w,1/2h}{v4a}
\fmfforce{5.5/6w,1/2h}{v4b}
\fmf{fermion}{v3,v1}
\fmf{fermion}{v2,v3}
\fmf{plain,left=1}{v4b,v4a}
\fmf{fermion,right=1}{v4b,v4a}
\fmfdot{v3}
\fmfv{decor.size=0, label=${\scriptstyle 1}$, l.dist=1mm, l.angle=0}{v2}
\fmfv{decor.size=0, label=${\scriptstyle 4}$, l.dist=1mm, l.angle=180}{v1}
\end{fmfgraph*}  \end{center}}
\hspace*{0.2cm} + 4 \hspace*{0.2cm}
\setlength{\unitlength}{1mm}
\parbox{14mm}{\begin{center}
\begin{fmfgraph*}(10,5)
\setval
\fmfstraight
\fmfforce{0w,0h}{i1}
\fmfforce{0w,1h}{i2}
\fmfforce{1w,0h}{o1}
\fmfforce{1w,1h}{o2}
\fmfforce{2.5/10w,0.5h}{v1}
\fmfforce{7.5/10w,0.5h}{v2}
\fmf{fermion}{i1,v1}
\fmf{fermion}{o1,v2}
\fmf{fermion}{v1,i2}
\fmf{fermion}{v2,o2}
\fmf{fermion,left=1}{v1,v2}
\fmf{fermion,left=1}{v2,v1}
\fmfdot{v1,v2}
\fmfv{decor.size=0, label=${\scriptstyle 1}$, l.dist=1mm, l.angle=180}{i1}
\fmfv{decor.size=0, label=${\scriptstyle 2}$, l.dist=1mm, l.angle=180}{i2}
\fmfv{decor.size=0, label=${\scriptstyle 3}$, l.dist=1mm, l.angle=0}{o1}
\fmfv{decor.size=0, label=${\scriptstyle 4}$, l.dist=1mm, l.angle=0}{o2}
\end{fmfgraph*}
\end{center}}
\hspace*{0.2cm} + 4 \hspace*{0.2cm} 
\setlength{\unitlength}{1mm}
\parbox{14mm}{\begin{center}
\begin{fmfgraph*}(10,5)
\setval
\fmfstraight
\fmfforce{0w,0h}{i1}
\fmfforce{0w,1h}{i2}
\fmfforce{1w,0h}{o1}
\fmfforce{1w,1h}{o2}
\fmfforce{2.5/10w,0.5h}{v1}
\fmfforce{7.5/10w,0.5h}{v2}
\fmf{fermion}{i1,v1}
\fmf{fermion}{o1,v2}
\fmf{fermion}{v1,i2}
\fmf{fermion}{v2,o2}
\fmf{fermion,left=1}{v1,v2}
\fmf{fermion,left=1}{v2,v1}
\fmfdot{v1,v2}
\fmfv{decor.size=0, label=${\scriptstyle 3}$, l.dist=1mm, l.angle=180}{i1}
\fmfv{decor.size=0, label=${\scriptstyle 2}$, l.dist=1mm, l.angle=180}{i2}
\fmfv{decor.size=0, label=${\scriptstyle 1}$, l.dist=1mm, l.angle=0}{o1}
\fmfv{decor.size=0, label=${\scriptstyle 4}$, l.dist=1mm, l.angle=0}{o2}
\end{fmfgraph*}
\end{center}}
\hspace*{0.2cm} + 2 \hspace*{0.2cm} 
\setlength{\unitlength}{1mm}
\parbox{14mm}{\begin{center}
\begin{fmfgraph*}(10,5)
\setval
\fmfstraight
\fmfforce{0w,0h}{i1}
\fmfforce{0w,1h}{i2}
\fmfforce{1w,0h}{o1}
\fmfforce{1w,1h}{o2}
\fmfforce{2.5/10w,0.5h}{v1}
\fmfforce{7.5/10w,0.5h}{v2}
\fmf{fermion}{v1,i1}
\fmf{fermion}{o1,v2}
\fmf{fermion}{v1,i2}
\fmf{fermion}{o2,v2}
\fmf{fermion,right=1}{v2,v1}
\fmf{fermion,left=1}{v2,v1}
\fmfdot{v1,v2}
\fmfv{decor.size=0, label=${\scriptstyle 4}$, l.dist=1mm, l.angle=180}{i1}
\fmfv{decor.size=0, label=${\scriptstyle 2}$, l.dist=1mm, l.angle=180}{i2}
\fmfv{decor.size=0, label=${\scriptstyle 1}$, l.dist=1mm, l.angle=0}{o1}
\fmfv{decor.size=0, label=${\scriptstyle 3}$, l.dist=1mm, l.angle=0}{o2}
\end{fmfgraph*}
\end{center}}
\end{eqnarray}
leads with $p=2$ in Eq. (\ref{HUGOREK}) to the third-order contribution in Table \ref{resu} which also shows the results
for the subsequent order. We remark that the grand-canonical
potential for the dilute bose gas has been calculated so far up to three loops in the seminal paper \cite{Yang}.
Recently, this result has been used to determine the critical temperature for the homogeneous dilute bose gas up to second order
\cite{Arnold}.
\end{fmffile}
\section*{Acknowledgement}
We thank Michael Bachmann, Hagen Kleinert, and Jens Koch for discussions and for a critical reading of the manuscript. One of us,
A.P., is grateful to Roman Jackiw and Kerson Huang for their kind hospitality at the Center for Theoretical Physics at MIT
where this work was completed. 
Both of us cordially congratulate Jozef Devreese to his 65th birthday.
\setlength{\unitlength}{1mm}
\newpage
\begin{fmffile}{hugo2}
\begin{table}[t]
\begin{center}
\begin{tabular}{|c|c|}
\hline\hline  
\,\,\,$p$\,\,\,
&
\begin{tabular}{@{}c}
$\mbox{}$ \\*[4mm]
$\mbox{}$
\end{tabular}
  \parbox{10mm}{\centerline{
  \begin{fmfgraph*}(8,6)
  \setval
  \fmfforce{0w,1/2h}{v1}
  \fmfforce{1w,1/2h}{v2}
  \fmfforce{1/2w,1h}{v3}
  \fmfforce{1/2w,0h}{v4}
  \fmfforce{1/2w,1/2h}{v5}
  \fmf{plain,left=0.4}{v1,v3,v2,v4,v1}
  \fmfv{decor.size=0, label=${\scriptstyle p}$, l.dist=0mm, l.angle=0}{v5}
  \end{fmfgraph*} } }  
\\
\hline
$1$ &
\hspace{-10pt}
\rule[-10pt]{0pt}{26pt}
  \begin{tabular}{@{}c}
  $\mbox{}$\\
  ${\scriptstyle \mbox{\#1.1}}$ \\
  $1$ \\ ${\scriptstyle ( 0, 0 , 0 ; 1 )}$\\
  $\mbox{}$
  \end{tabular}
\parbox{8mm}{\begin{center}
\begin{fmfgraph}(10,5)
\setval
\fmfforce{1/4w,1h}{o1}
\fmfforce{1/4w,0h}{u1}
\fmfforce{1/2w,1/2h}{v2}
\fmfforce{3/4w,1h}{o3}
\fmfforce{3/4w,0h}{u3}
\fmf{fermion,right=1}{o1,u1}
\fmf{fermion,right=1}{u3,o3}
\fmf{plain,right=1}{o3,u3}
\fmf{plain,right=1}{u1,o1}
\fmfdot{v2}
\end{fmfgraph}\end{center}}
\\
\hline
$2$ &
\hspace{-10pt}
\rule[-10pt]{0pt}{26pt}
  \begin{tabular}{@{}c}
  $\mbox{}$\\
  ${\scriptstyle \mbox{\#2.1}}$ \\
  $1/2$ \\ 
  ${\scriptstyle ( 0, 0 , 1 ; 2 )}$\\
  $\mbox{}$
  \end{tabular}
\parbox{7.5mm}{\begin{center}
\begin{fmfgraph}(7.5,7.5)
\setval
\fmfforce{0w,0.5h}{v1}
\fmfforce{1w,0.5h}{v2}
\fmf{fermion,right=1}{v1,v2,v1}
\fmf{fermion,left=0.4}{v1,v2,v1}
\fmfdot{v1,v2}
\end{fmfgraph}\end{center}} 
\hspace*{5mm}
  \begin{tabular}{@{}c}
  ${\scriptstyle \mbox{\#2.2}}$ \\
  $2$ \\ 
  ${\scriptstyle ( 0, 0 , 0 ; 2 )}$
  \end{tabular}
\parbox{15mm}{\begin{center}
\begin{fmfgraph}(15,5)
\setval
\fmfforce{1/6w,1h}{o1}
\fmfforce{1/6w,0h}{u1}
\fmfforce{1/3w,1/2h}{v2}
\fmfforce{2/3w,1/2h}{v4}
\fmfforce{5/6w,1h}{o3}
\fmfforce{5/6w,0h}{u3}
\fmf{fermion,right=1}{o1,u1}
\fmf{fermion,right=1}{u3,o3}
\fmf{plain,right=1}{o3,u3}
\fmf{plain,right=1}{u1,o1}
\fmf{fermion,right=1}{v4,v2,v4}
\fmfdot{v2,v4}
\end{fmfgraph}\end{center}}
\\
\hline
$3$ &
\hspace*{1mm}
  \begin{tabular}{@{}c}
  $\mbox{}$\\
  ${\scriptstyle \mbox{\#3.1}}$ \\
  $4/3$ \\ ${\scriptstyle ( 0, 0 , 0 ; 6 )}$ \\
  $\mbox{}$
  \end{tabular}
\parbox{6mm}{\begin{center}
\begin{fmfgraph}(8,8)
\setval
\fmfforce{2/8w,0h}{v1}
\fmfforce{2/8w,1h}{v2}
\fmfforce{-1.5/8w,0.25h}{v3}
\fmfforce{5.5/8w,0.25h}{v4}
\fmf{fermion,right=0.55}{v2,v3,v4,v2}
\fmf{fermion}{v2,v4,v3,v2}
\fmfdot{v2,v3,v4}
\end{fmfgraph}
\end{center}} 
\hspace*{3mm}
  \begin{tabular}{@{}c}
  $\mbox{}$\\
  ${\scriptstyle \mbox{\#3.2}}$ \\
  $1/3$ \\ ${\scriptstyle ( 3, 0 , 0 ; 3 )}$ \\
  $\mbox{}$
  \end{tabular}
\parbox{8mm}{\begin{center}
\begin{fmfgraph}(8,8)
\setval
\fmfforce{2/8w,0h}{v1}
\fmfforce{2/8w,1h}{v2}
\fmfforce{-1.5/8w,0.25h}{v3}
\fmfforce{5.5/8w,0.25h}{v4}
\fmf{fermion,right=0.55}{v2,v3,v4,v2}
\fmf{fermion}{v2,v3,v4,v2}
\fmfdot{v2,v3,v4}
\end{fmfgraph}
\end{center}} 
\hspace{1mm}
  \begin{tabular}{@{}c}
  ${\scriptstyle \mbox{\#3.3}}$ \\
  $4$\\ 
  ${\scriptstyle ( 0, 1 , 0 ; 1 )}$
  \end{tabular}
\parbox{6.5mm}{\begin{center}
\begin{fmfgraph}(7.5,12.5)
\setval
\fmfforce{0w,0.3h}{v1}
\fmfforce{1w,0.3h}{v2}
\fmfforce{0.5w,0.6h}{v3}
\fmfforce{1.25/7.5w,10/12.5h}{i4}
\fmfforce{6.25/7.5w,10/12.5h}{o4}
\fmf{fermion,right=1}{v1,v2}
\fmf{fermion,right=0.4}{v2,v3,v1}
\fmf{fermion,left=0.4}{v1,v2,v1}
\fmf{fermion,right=1}{o4,i4}
\fmf{plain,right=1}{i4,o4}
\fmfdot{v1,v2,v3}
\end{fmfgraph}\end{center}} 
\quad
  \begin{tabular}{@{}c}
  ${\scriptstyle \mbox{\#3.4}}$ \\
  $8/3$\\ 
  ${\scriptstyle ( 0, 0 , 0 ; 3 )}$
  \end{tabular}
\parbox{12mm}{\begin{center}
\begin{fmfgraph}(13,13)
\setval
\fmfforce{1/2w,3/13h}{v1}
\fmfforce{1/2w,8/13h}{v2}
\fmfforce{1/2w,1h}{v3}
\fmfforce{4/13w,10.5/13h}{i3}
\fmfforce{9/13w,10.5/13h}{o3}
\fmfforce{4.25/13w,4.25/13h}{v4}
\fmfforce{0.75/13w,5/13h}{i4}
\fmfforce{3.75/13w,1/13h}{o4}
\fmfforce{8.75/13w,4.25/13h}{v5}
\fmfforce{12.25/13w,5/13h}{i5}
\fmfforce{9.25/13w,1/13h}{o5}
\fmf{fermion,right=0.6}{v4,v5,v2,v4}
\fmf{fermion,right=1}{o3,i3}
\fmf{plain,right=1}{i3,o3}
\fmf{fermion,right=1}{i4,o4}
\fmf{plain,right=1}{o4,i4}
\fmf{fermion,right=1}{o5,i5}
\fmf{plain,right=1}{i5,o5}
\fmfdot{v2,v4,v5}
\end{fmfgraph}\end{center}} 
\quad
  \begin{tabular}{@{}c}
  ${\scriptstyle \mbox{\#3.5}}$ \\
  $4$\\ 
  ${\scriptstyle ( 0, 0 , 0 ; 2 )}$
  \end{tabular}
\parbox{19mm}{\begin{center}
\begin{fmfgraph}(20,5)
\setval
\fmfforce{0w,1h}{o1}
\fmfforce{0w,0h}{u1}
\fmfforce{2.5/20w,1/2h}{v2}
\fmfforce{7.5/20w,1/2h}{v4}
\fmfforce{12.5/20w,1/2h}{v6}
\fmfforce{15/20w,1h}{o3}
\fmfforce{15/20w,0h}{u3}
\fmf{fermion,right=1}{o1,u1}
\fmf{fermion,right=1}{u3,o3}
\fmf{plain,right=1}{o3,u3}
\fmf{plain,right=1}{u1,o1}
\fmf{fermion,right=1}{v4,v2,v4}
\fmf{fermion,right=1}{v6,v4,v6}
\fmfdot{v2,v4,v6}
\end{fmfgraph}\end{center}}
\\
\hline
& \\
&
  \begin{tabular}{@{}c}
  ${\scriptstyle \mbox{\#4.1}}$ \\
  $2$ \\ ${\scriptstyle ( 0 , 0 , 0 ; 8 )}$ 
  \end{tabular} 
\parbox{10mm}{\begin{center}
\begin{fmfgraph*}(10,10)
\setval
\fmfforce{0.1464466w,0.1464466h}{v1}
\fmfforce{0.1464466w,0.8535534h}{v2}
\fmfforce{0.8535534w,0.8535534h}{v3}
\fmfforce{0.8535534w,0.1464466h}{v4}
\fmfforce{1/2w,0h}{v5}
\fmfforce{1/2w,1h}{v6}
\fmf{fermion,right=0.4}{v1,v4,v3,v2,v1}
\fmf{fermion,right=0.1}{v1,v2,v3,v4,v1}
\fmfdot{v1,v2,v3,v4}
\end{fmfgraph*} \end{center}}
\hspace*{3mm}
  \begin{tabular}{@{}c}
  ${\scriptstyle \mbox{\#4.2}}$ \\
  $1/4$ \\ ${\scriptstyle ( 4 , 0 , 0 ; 4 )}$ 
  \end{tabular} 
\parbox{10mm}{\begin{center}
\begin{fmfgraph*}(10,10)
\setval
\fmfforce{0.1464466w,0.1464466h}{v1}
\fmfforce{0.1464466w,0.8535534h}{v2}
\fmfforce{0.8535534w,0.8535534h}{v3}
\fmfforce{0.8535534w,0.1464466h}{v4}
\fmfforce{1/2w,0h}{v5}
\fmfforce{1/2w,1h}{v6}
\fmf{fermion,right=0.4}{v1,v4,v3,v2,v1}
\fmf{fermion,left=0.1}{v1,v4,v3,v2,v1}
\fmfdot{v1,v2,v3,v4}
\end{fmfgraph*} \end{center}}
\quad
  \begin{tabular}{@{}c}
  ${\scriptstyle \mbox{\#4.3}}$ \\
  $4$ \\ 
  ${\scriptstyle ( 0 , 0 , 0 ; 4 )}$
  \end{tabular}
\parbox{10mm}{\begin{center}
\begin{fmfgraph*}(10,10)
\setval
\fmfforce{0w,1/2h}{v1}
\fmfforce{1w,1/2h}{v2}
\fmfforce{1/2w,1/4h}{v3}
\fmfforce{1/2w,3/4h}{v4}
\fmf{fermion,right=1}{v1,v2,v1}
\fmf{fermion,right=1}{v3,v4,v3}
\fmf{fermion,left=0.2}{v1,v4,v2,v3,v1}
\fmfdot{v1,v2,v3,v4}
\end{fmfgraph*} \end{center}}
\quad
  \begin{tabular}{@{}c}
  ${\scriptstyle \mbox{\#4.4}}$ \\
  $4$ \\ 
  ${\scriptstyle ( 1 , 0 , 0 ; 2 )}$
  \end{tabular}
\parbox{10mm}{\begin{center}
\begin{fmfgraph*}(10,10)
\setval
\fmfforce{0w,1/2h}{v1}
\fmfforce{1w,1/2h}{v2}
\fmfforce{1/2w,1/4h}{v3}
\fmfforce{1/2w,3/4h}{v4}
\fmf{fermion,right=1}{v1,v2,v1}
\fmf{fermion,left=1}{v3,v4}
\fmf{fermion,right=1}{v3,v4}
\fmf{fermion,right=0.2}{v4,v1}
\fmf{fermion,left=0.2}{v4,v2}
\fmf{fermion,right=0.2}{v3,v2}
\fmf{fermion,left=0.2}{v3,v1}
\fmfdot{v1,v2,v3,v4}
\end{fmfgraph*} \end{center}}
\quad
\\
  $4$ & \begin{tabular}{@{}c}
  ${\scriptstyle \mbox{\#4.5}}$ \\
  $16$ \\ 
  ${\scriptstyle (  0 , 0 , 0 ; 1 )}$
  \end{tabular}
\parbox{8mm}{\begin{center}
\begin{fmfgraph*}(8,13)
\setval
\fmfforce{0.5w,0h}{v1}
\fmfforce{0.5w,8/13h}{v2}
\fmfforce{0.0669873w,0.46154h}{v3}
\fmfforce{0.933w,0.46154h}{v4}
\fmfforce{0.5w,1h}{v5}
\fmfforce{1.5/8w,10.5/13h}{i5}
\fmfforce{6.5/8w,10.5/13h}{o5}
\fmf{fermion,right=0.25}{v4,v2,v3}
\fmf{fermion,right=0.55}{v3,v1,v4}
\fmf{fermion,right=1}{o5,i5}
\fmf{plain,right=1}{i5,o5}
\fmf{fermion}{v3,v4,v1,v3}
\fmfdot{v1,v2,v3,v4}
\end{fmfgraph*}\end{center}}
\quad
  \begin{tabular}{@{}c}
  ${\scriptstyle \mbox{\#4.6}}$ \\
  $4$ \\ 
  ${\scriptstyle ( 2  , 0 , 0 ; 1 )}$
  \end{tabular}
\parbox{8mm}{\begin{center}
\begin{fmfgraph*}(8,13)
\setval
\fmfforce{0.5w,0h}{v1}
\fmfforce{0.5w,8/13h}{v2}
\fmfforce{0.0669873w,0.46154h}{v3}
\fmfforce{0.933w,0.46154h}{v4}
\fmfforce{1.5/8w,10.5/13h}{i5}
\fmfforce{6.5/8w,10.5/13h}{o5}
\fmf{fermion,right=0.25}{v4,v2,v3}
\fmf{fermion,right=0.55}{v3,v1,v4}
\fmf{fermion,right=1}{o5,i5}
\fmf{plain,right=1}{i5,o5}
\fmf{fermion}{v4,v3,v1,v4}
\fmfdot{v1,v2,v3,v4}
\end{fmfgraph*}\end{center}}
\quad
  \begin{tabular}{@{}c}
  ${\scriptstyle \mbox{\#4.7}}$ \\
  $8$ \\ 
  ${\scriptstyle ( 0, 0 , 0 ; 2 )}$
  \end{tabular}
\parbox{7.5mm}{\begin{center}
\begin{fmfgraph*}(7.5,17.5)
\setval
\fmfforce{0w,0.5h}{v1}
\fmfforce{1w,0.5h}{v2}
\fmfforce{0.5w,h}{v3}
\fmfforce{0.5w,1h}{v4}
\fmfforce{0.5w,0h}{v5}
\fmfforce{0.5w,0.2857h}{v6}
\fmfforce{0.5w,0.71429h}{v7}
\fmfforce{1.25/7.5w,15/17.5h}{i7}
\fmfforce{6.25/7.5w,15/17.5h}{o7}
\fmfforce{1.25/7.5w,2.5/17.5h}{i6}
\fmfforce{6.25/7.5w,2.5/17.5h}{o6}
\fmf{fermion,right=1}{o7,i7}
\fmf{plain,right=1}{i7,o7}
\fmf{fermion,right=1}{i6,o6}
\fmf{plain,right=1}{o6,i6}
\fmf{fermion,right=0.4}{v1,v6,v2,v7,v1}
\fmf{fermion,left=0.4}{v1,v2,v1}
\fmfdot{v1,v2,v6,v7}
\end{fmfgraph*}\end{center}}
\quad
  \begin{tabular}{@{}c}
  ${\scriptstyle \mbox{\#4.8}}$ \\
  $4$ \\ 
  ${\scriptstyle ( 1, 0 , 0 ; 2 )}$
  \end{tabular}
\parbox{7.5mm}{\begin{center}
\begin{fmfgraph*}(7.5,17.5)
\setval
\fmfforce{0w,0.5h}{v1}
\fmfforce{1w,0.5h}{v2}
\fmfforce{0.5w,h}{v3}
\fmfforce{0.5w,1h}{v4}
\fmfforce{0.5w,0h}{v5}
\fmfforce{0.5w,0.2857h}{v6}
\fmfforce{0.5w,0.71429h}{v7}
\fmfforce{1.25/7.5w,15/17.5h}{i7}
\fmfforce{6.25/7.5w,15/17.5h}{o7}
\fmfforce{1.25/7.5w,2.5/17.5h}{i6}
\fmfforce{6.25/7.5w,2.5/17.5h}{o6}
\fmf{fermion,right=1}{o7,i7}
\fmf{plain,right=1}{i7,o7}
\fmf{fermion,left=1}{o6,i6}
\fmf{plain,right=1}{o6,i6}
\fmf{fermion,right=0.4}{v2,v7,v1,v2}
\fmf{fermion,left=0.4}{v2,v6,v1,v2}
\fmfdot{v1,v2,v6,v7}
\end{fmfgraph*}\end{center}} 
\quad
  \begin{tabular}{@{}c}
  ${\scriptstyle \mbox{\#4.9}}$ \\
  $2$ \\ 
  ${\scriptstyle ( 0, 2, 0; 2 )}$
  \end{tabular}
\parbox{10mm}{\begin{center}
\begin{fmfgraph*}(10,12.5)
\setval
\fmfforce{1/4w,1/5h}{v1}
\fmfforce{3/4w,1/5h}{v2}
\fmfforce{1/4w,4/5h}{v3}
\fmfforce{3/4w,4/5h}{v4}
\fmf{fermion}{v1,v2}
\fmf{fermion,right=1}{v1,v2,v1}
\fmf{fermion}{v4,v3}
\fmf{fermion,right=1}{v3,v4,v3}
\fmf{fermion,right=0.5}{v3,v1}
\fmf{fermion,right=0.5}{v2,v4}
\fmfdot{v1,v2,v3,v4}
\end{fmfgraph*} \end{center}}
\quad
\\
  $ $ & \begin{tabular}{@{}c}
  ${\scriptstyle \mbox{\#4.10}}$ \\
  $8$ \\ 
  ${\scriptstyle (  0 , 1 , 0 ; 1 )}$
  \end{tabular}
\parbox{12mm}{\begin{center}
\begin{fmfgraph*}(13,13)
\setval
\fmfforce{2.5/13w,0.3h}{v1}
\fmfforce{10.5/13w,0.3h}{v2}
\fmfforce{4.1/13w,0.55h}{v3}
\fmfforce{8.9/13w,0.55h}{v4}
\fmfforce{0w,0.7h}{i5}
\fmfforce{5/13w,0.7h}{o5}
\fmfforce{8/13w,0.7h}{i6}
\fmfforce{1w,0.7h}{o6}
\fmf{fermion,right=1}{v1,v2}
\fmf{fermion,right=0.3}{v2,v4,v3,v1}
\fmf{fermion,left=0.4}{v1,v2,v1}
\fmf{fermion,right=1}{o5,i5}
\fmf{plain,right=1}{i5,o5}
\fmf{fermion,right=1}{o6,i6}
\fmf{plain,right=1}{i6,o6}
\fmfdot{v1,v2,v3,v4}
\end{fmfgraph*}\end{center}}
\quad
  \begin{tabular}{@{}c}
  ${\scriptstyle \mbox{\#4.11}}$ \\
  $8$ \\ 
  ${\scriptstyle ( 0, 1 , 0 ; 1 )}$
  \end{tabular}
\parbox{7.5mm}{\begin{center}
\begin{fmfgraph*}(7.5,17.5)
\setval
\fmfforce{0w,0.21429h}{v1}
\fmfforce{1w,0.21429h}{v2}
\fmfforce{0.5w,0.42857h}{v3}
\fmfforce{0.5w,0.71429h}{v4}
\fmfforce{1.25/7.5w,15/17.5h}{i5}
\fmfforce{6.25/7.5w,15/17.5h}{o5}
\fmf{fermion,right=1}{o5,i5}
\fmf{plain,right=1}{i5,o5}
\fmf{fermion,right=1}{v1,v2}
\fmf{fermion,right=0.4}{v2,v3,v1}
\fmf{fermion,right=1}{v3,v4,v3}
\fmf{fermion,left=0.4}{v1,v2,v1}
\fmfdot{v1,v2,v3,v4}
\end{fmfgraph*}\end{center}} 
\quad
  \begin{tabular}{@{}c}
  ${\scriptstyle \mbox{\#4.12}}$ \\
  $16$ \\ 
  ${\scriptstyle ( 0, 0, 0; 1 )}$
  \end{tabular}
\parbox{15mm}{\begin{center}
\begin{fmfgraph*}(13,18)
\setval
\fmfforce{1/2w,3/18h}{v1}
\fmfforce{1/2w,8/18h}{v2}
\fmfforce{1/2w,13/18h}{v3}
\fmfforce{4/13w,15.5/18h}{i3}
\fmfforce{9/13w,15.5/18h}{o3}
\fmfforce{4.25/13w,4.25/18h}{v4}
\fmfforce{0.75/13w,5/18h}{i4}
\fmfforce{3.75/13w,1/18h}{o4}
\fmfforce{8.75/13w,4.25/18h}{v5}
\fmfforce{12.25/13w,5/18h}{i5}
\fmfforce{9.25/13w,1/18h}{o5}
\fmf{fermion,right=0.6}{v4,v5,v2,v4}
\fmf{fermion,right=1}{o3,i3}
\fmf{plain,right=1}{i3,o3}
\fmf{fermion,right=1}{i4,o4}
\fmf{plain,right=1}{o4,i4}
\fmf{fermion,right=1}{o5,i5}
\fmf{plain,right=1}{i5,o5}
\fmf{fermion,right=1}{v2,v3,v2}
\fmfdot{v2,v3,v4,v5}
\end{fmfgraph*}\end{center}} 
\quad
\\
& 
  \begin{tabular}{@{}c}
  ${\scriptstyle \mbox{\#4.13}}$ \\
  $4$ \\ 
  ${\scriptstyle ( 0, 0, 0; 4 )}$ 
  \end{tabular}
\parbox{15mm}{\begin{center}
\begin{fmfgraph*}(15,15)
\setval
\fmfforce{1/3w,1/2h}{v1}
\fmfforce{2/3w,1/2h}{v2}
\fmfforce{1/2w,1/3h}{v3}
\fmfforce{1/2w,2/3h}{v4}
\fmfforce{5/15w,12.5/15h}{i4}
\fmfforce{10/15w,12.5/15h}{o4}
\fmf{fermion,right=1}{o4,i4}
\fmf{plain,right=1}{i4,o4}
\fmfforce{5/15w,2.5/15h}{i3}
\fmfforce{10/15w,2.5/15h}{o3}
\fmf{fermion,right=1}{i3,o3}
\fmf{plain,right=1}{o3,i3}
\fmfforce{2.5/15w,5/15h}{i1}
\fmfforce{2.5/15w,10/15h}{o1}
\fmf{fermion,right=1}{o1,i1}
\fmf{plain,right=1}{i1,o1}
\fmfforce{12.5/15w,5/15h}{i2}
\fmfforce{12.5/15w,10/15h}{o2}
\fmf{fermion,right=1}{i2,o2}
\fmf{plain,right=1}{o2,i2}
\fmf{fermion,right=0.4}{v1,v3,v2,v4,v1}
\fmfdot{v1,v2,v3,v4}
\end{fmfgraph*}\end{center}} 
\hspace*{5mm}
  \begin{tabular}{@{}c}
  ${\scriptstyle \mbox{\#4.14}}$ \\
  $8$ \\ 
  ${\scriptstyle ( 0, 0, 0; 2 )}$ 
  \end{tabular}
\parbox{25mm}{\begin{center}
\begin{fmfgraph*}(25,5)
\setval
\fmfforce{1/10w,1h}{o1}
\fmfforce{1/10w,0h}{u1}
\fmfforce{1/5w,1/2h}{v2}
\fmfforce{2/5w,1/2h}{v4}
\fmfforce{3/5w,1/2h}{v5}
\fmfforce{4/5w,1/2h}{v6}
\fmfforce{9/10w,1h}{o3}
\fmfforce{9/10w,0h}{u3}
\fmf{fermion,right=1}{o1,u1}
\fmf{fermion,right=1}{u3,o3}
\fmf{plain,right=1}{o3,u3}
\fmf{plain,right=1}{u1,o1}
\fmf{fermion,right=1}{v4,v2,v4}
\fmf{fermion,right=1}{v4,v5,v4}
\fmf{fermion,right=1}{v6,v5,v6}
\fmfdot{v2,v4,v5,v6}
\end{fmfgraph*} \end{center}}
\\
& \\  
\hline\hline
\end{tabular}
\end{center}
\caption{\label{resu}
Hugenholtz diagrams and their weights for the grand-canonical potential of a weakly interacting bose gas.
Each diagram is characterized by the vector $(D,T,F;N)$ whose components lead to the weight according to Eq. (\ref{WEI}).} 
\end{table}
\end{fmffile}
\end{document}